\documentclass[a4paper,fleqn,usenatbib]{mnras}

\usepackage[T1]{fontenc}
\usepackage{ae,aecompl}

\usepackage{graphics,epsfig,psfig}
\usepackage{dblfloatfix}
\usepackage[normalem]{ulem}
\usepackage[dvipsnames]{color}
\usepackage[]{inputenc,amssymb}
\usepackage{multirow, multicol}
\usepackage{booktabs,caption}
\usepackage{lipsum}
\usepackage{widetext}
\usepackage{pifont}
\usepackage{natbib}
\usepackage{dblfloatfix,afterpage}
\usepackage[normalem]{ulem}
\usepackage[table]{xcolor}
\usepackage{graphicx}	
\usepackage{amsmath}	
\usepackage{amssymb}	
\usepackage{array,multirow}

\newcommand{\BLACK}{\color{black}}
\newcommand{\BLUE}{\color{blue}}

\newcommand{\quotes}[1]{``#1''}
\newcommand{\pd}[2]{\frac{\partial #1}{\partial #2} } 
\newcommand{\DS}{\displaystyle}

\newcommand{\cL}{{\cal L}}
\newcommand{\cS}{{\cal S}}
\newcommand{\cU}{{\cal U}}
\newcommand{\cV}{{\cal V}}
\newcommand{\tens}[1]{\mathsf{#1}}
\newcommand{\HALF}{\frac{1}{2}}
\newcommand{\PLUTO}{{\bf \textsf{\small PLUTO}} }

\definecolor{dg}{rgb}{0.0, 0.5, 0.0}
\newcommand{\GREEN}{\color{dg}}

\title[Implementation of cosmic ray two-fluid equations]{
A numerical approach to the non-uniqueness problem of cosmic ray two-fluid equations at shocks}
\author[Gupta, Sharma, \& Mignone]{
Siddhartha Gupta$^{1,2,3}$\thanks{E-mail: gsiddhartha@uchicago.edu},
Prateek Sharma$^{1}$,
Andrea Mignone$^{4}$\\
$^{1}$Department of Physics, Indian Institute of Science, Bangalore 560012, India\\
$^{2}$Raman Research Institute, Sadashiva Nagar, Bangalore 560080, India\\
$^{3}$Department of Astronomy and Astrophysics, University of Chicago, IL 60637, USA\\
$^{4}$Dipartimento di Fisica, Universit$\grave{a}$ di Torino, via P. Giuria 1, I-10125 Torino, Italy
}
\date{\null}
\begin{document}
\label{firstpage}
\pagerange{\pageref{firstpage}--\pageref{lastpage}}
\maketitle
\begin{abstract}
Cosmic rays (CRs) are frequently modeled as an additional fluid in hydrodynamic (HD) and magnetohydrodynamic (MHD) simulations of astrophysical flows. The standard CR two-fluid model is described in terms of three conservation laws (expressing conservation of mass, momentum and total energy) and one additional equation (for the CR pressure) that cannot be cast in a satisfactory conservative form. The presence of non-conservative terms with spatial derivatives in the model equations prevents a unique weak solution behind a shock. We investigate a number of methods for the numerical solution of the two-fluid equations and find that, in the presence of shock waves, the results generally depend on the numerical details (spatial reconstruction, time stepping, the CFL number, and the adopted discretization). All methods converge to a unique result if the energy partition between the thermal and non-thermal fluids at the shock is prescribed using a subgrid prescription. This highlights the non-uniqueness problem of the two-fluid equations at shocks. From our numerical investigations, we report a robust method for which the solutions are insensitive to the numerical details even in absence of a subgrid prescription, although we recommend a subgrid closure at shocks using results from kinetic theory. The subgrid closure is crucial for a reliable post-shock solution and also its impact on large scale flows because the shock microphysics that determines CR acceleration is not accurately captured in a fluid approximation. Critical test problems, limitations of fluid modeling, and future directions are discussed.
\end{abstract}
\BLACK
\begin{keywords}
shock waves  -- cosmic rays -- hydrodynamics -- methods:numerical
\end{keywords}
\section{Introduction}

Macroscopic extension of the cosmic ray (CR) transport equation in the form of the two-fluid CR-HD/MHD equations (\citealt{Skilling1971}; \citealt{Drury1981}) is very convenient to study the effects of non-thermal processes in astrophysical systems. 
A two fluid-model provides important insights about the dynamical effects of CRs on large scales (such as blast waves, wind bubbles, galaxies; see e.g., \citealt{Chevalier1983,Salem2014,Gupta2018a}), which are expensive to capture from the CR kinetic theory (see \citealt{Drury1983} for a review). In many astrophysical systems, the CR energy density is comparable to the thermal/magnetic energy density, and the CR pressure cannot be ignored.

A fluid description of CRs is justified since the Larmor radius of energy density-dominating CRs ($\sim 10^{-5}\,E_{\rm 10GeV}/B_{\rm \mu G}\,{\rm pc}$) is much smaller than the scales of interest and they are confined along the direction of magnetic fields by self-generated magnetic fluctuations at this scale (\citealt{Kulsrud1969}). Therefore, the two-fluid model is applicable in a variety of astrophysical systems, ranging from a star-forming cloud to clusters of galaxies.

Many astrophysical phenomena on large scales are studied with the two-fluid model of CRs. CRs represent an attractive agent for feedback heating as their energy loss time scale is much longer than the cooling time of the thermal gas. CRs retain energy in the cloud for a long time and provide extra pressure leading to a moderate star formation rate. For a similar reason, CRs help to launch galactic winds (e.g., \citealt{Salem2014}; \citealt{Wiener2017}). Investigations with two-fluid models also found that CRs can reduce the temperature of the circumgalactic medium and account for the observed absorption lines of various elements (e.g., \citealt{Butsky2018}). CR heating can be an efficient heating mechanism in galaxy clusters (e.g., \citealt{Guo2008}) and can affect buoyancy instabilities in the intracluster medium (e.g., \citealt{Sharma2009}). These important conclusions are based on numerical simulations of the two-fluid model and therefore it is necessary to closely examine the properties of these equations and their numerical implementation.

The two-fluid CR-HD model contains three conservation laws (for mass, momentum, and total energy) plus one additional equation, which governs the evolution of CR energy density. Since the CR particle density is usually negligible compared to the thermal particle density, the evolution of the CR particle density is not considered. However, a CR particle carries significant kinetic energy, which makes the energy density contribution of CRs (integrated over entire spectrum) comparable to that of the thermal plasma.

Similar two-fluid equations are also used in other contexts. For example, the two-fluid MHD equations (with pressures along and perpendicular to field lines and equations governing their evolution, and the generalization to include electrons and ions separately) are used to study pressure anisotropy in astrophysical and fusion plasmas (e.g., \citealt{Sharma2006}; \citealt{Jardin2007}). Both CR-HD/MHD and MHD with anisotropic pressure have two internal energy equations but they do not have a separate density/velocity equation for the second fluid. In the CR two-fluid model, the equation that determines the evolution of CR energy density cannot be cast in a conservative form owing to the presence of a source (or coupling) term containing spatial derivatives (either $p_{\rm cr}\nabla\cdot{\bf v}$ or ${\bf v}\cdot\nabla p_{\rm cr}$). While this term does not represent an issue for smooth flows, it poses a serious challenge in determining the weak form of the equations. Therefore, the shock jump conditions are non-unique and the solutions across a shock depend on numerical details. Note that there are other two-fluid systems that do not have the spatial derivatives in the source terms and for these the solution across a shock is unique (e.g., \citealt{Balsara2016}).

The numerical discretization of the non-conservative terms in the two-fluid CR-HD/MHD system can be done in different ways. However, these implementations may not produce an {\it identical} solution. This crucial point was highlighted by \citet{Kudoh2016}. They suggested that if the CR energy density is estimated from the advection of a passive scalar, namely $\chi = p^{1/\gamma_{\rm cr}}_{\rm cr}/\rho$ (where $\gamma_{\rm cr}$ is adiabatic index for CR fluid, $p_{\rm cr}$ is CR pressure, $\rho$ is gas density), then the results are consistent with the underlying mathematical formalism. This is equivalent to positing that the CR entropy $p_{\rm cr}/\rho^{\gamma_{\rm cr}}$ is conserved across a shock. 
One advantage of this method is that the two-fluid equations apparently become conservative, which makes numerical application of Godunov-type shock-capturing schemes straightforward. However, the assumption of a constant CR entropy across a shock is inconsistent with the idea that most CRs are accelerated at shocks. This motivates us to explore more physically motivated strategies.

In this paper, we discuss several numerical discretizations of the two-fluid CR-HD equations by implementing the equations in the \PLUTO code (\citealt{Mignone2007}). We find that for most of the common methods, the numerical solutions depend on the choice of spatial reconstruction, time stepping, and even the CFL number. We suggest a method which gives robust numerical results (i.e., solutions are less sensitive to these choices compared to other methods). A physically reliable two-fluid implementation of CRs must be calibrated with the results from kinetic (e.g., Particle-In-Cell) simulations. Our results are crucial for simulating the impact of CRs on large scales (e.g., galaxy), where the CR acceleration microphysics is difficult to resolve.

The non-uniqueness problem that we discuss in this paper applies to both the two-fluid hydro and MHD equations but we exclusively consider the former to focus on this problem and its possible solution in a simpler setting. Generalization to MHD is straightforward.

We organize this paper as follows. After presenting the basic equations in section \ref{sec:basics}, we discuss the closure problems in section \ref{subsec:closure}. Different methods for implementing two-fluid equations are given in section \ref{sec:method}. Section \ref{sec:result} presents results of various test problems. Section \ref{sec:discussions} discusses broader implications of our work. Our main results are summarized in section \ref{sec:conclu}.

\section{Governing Equations}\label{sec:basics}
The two-fluid CR HD/MHD equations are obtained from the Fokker-Planck CR transport equation (\citealt{Skilling1975}), which is given by
\begin{equation} \label{eq:crtransfull}
  \begin{array}{l}
  \DS \pd{f}{t} + \left({\bf v}+{\bf v}_{\rm st}\right)\cdot{\nabla}f =
   \frac{\mathcal{P}}{3}\pd{f}{\mathcal{P}} \nabla\cdot\left({\bf v}+{\bf v}_{\rm st}\right)
    \\ \noalign{\medskip}
   \DS + {\nabla} \cdot\left[D_{\rm s}\hat{b}\left(\hat{b}\cdot{\nabla}f\right)\right]
       + \frac{1}{\mathcal{P}^2}\pd{}{\mathcal{P}}\left(\mathcal{P}^{2}D_{\rm \mathcal{P}}
         \pd{f}{\mathcal{P}}\right).
  \end{array}
\end{equation}
Here $f({\bf x},{\bf \mathcal{P}},t)$ is the CR distribution function assumed to be isotropic in momentum space, $\mathcal{P}$ is the CR momentum, $D_{\rm s}$ and $D_{\rm \mathcal{P}}$ are the diffusion coefficients in spatial and momentum space, ${\bf v}$ is the velocity of thermal plasma. The term ${\bf v}_{\rm st} = (v_{\rm x}^{st}, v_{\rm y}^{st}, v_{\rm z}^{st})^\intercal$ represents the bulk velocity of the scattering centers of CR particles w.r.t. the background plasma, known as the streaming velocity, which is along the direction of the magnetic field ($\hat{b}$) but down the gradient of CR pressure (for a brief discussion see Appendix A in \citealt{Pfrommer2017}). The first term on the RHS represents the CR convection term, the second term is spatial diffusion while the third term represents CR diffusion in the momentum space (\citealt{Skilling1975}).

\citet{Drury1981} suggested that Eq. (\ref{eq:crtransfull}) can be simplified if one rewrites it in terms of macroscopic variables.
This is done by taking the energy moment of Eq. (\ref{eq:crtransfull}), which yields
\begin{eqnarray}
\frac{\partial\,e_{\rm cr}}{\partial t} +\left({\bf v}+{\bf v}_{\rm st}\right)\cdot{\nabla} e_{\rm cr} &=& \left(- e_{\rm cr}-p_{\rm cr}\right) {\nabla} \cdot\left({\bf v}+{\bf v}_{\rm st}\right) \nonumber \\
& & +{\nabla} \cdot\left[\kappa_{\rm cr}\hat{b}\left(\hat{b} \cdot {\nabla}e_{\rm cr}\right)\right] + \Gamma_{\rm m}\ .
\end{eqnarray}
Here $\kappa_{\rm cr}$ is the CR diffusion coefficient integrated over the CR distribution function (see equation $7$ in \citealt{Drury1981}).
CR streaming and anisotropic diffusion along field lines cannot be captured in hydrodynamics, therefore, we take $|{\bf v}|/|{\bf v}_{\rm st}|\gg 1$ and $\kappa_{\rm cr}\hat{b}(\hat{b} \cdot {\nabla}e_{\rm cr})\approx \kappa_{\rm cr}{\nabla}e_{\rm cr}$.
It is also assumed that CR diffusion in momentum space is negligible, so that $\Gamma_{\rm m}\rightarrow 0$.
This leads to the two-fluid CR-HD equations, written as
\begin{equation} \label{eq:gen}
 \pd{\cU}{t} + \nabla\cdot\tens{F}  =  \cS, \,
\end{equation}
where $\cU = \{\rho,\, \rho\,{\bf v},\, e_{\rm t},\, e_{\rm cr}\}^\intercal$ is the conservative variable array, while the flux tensor and the source term are given, respectively, by 
\begin{eqnarray} \label{eq:def}
 \tens{F} = \begin{bmatrix} 
\rho {\bf v}   \\
\\
\rho\,{\bf v} {\bf v} + p_{\rm t}\, {\tens{I}} \\
\\
\left( e_{\rm t}+ p_{\rm t}\right)\,{\bf v} - \kappa_{\rm cr} {\nabla}e_{\rm cr}\\
\\
 e_{\rm cr} \,{\bf v} -\kappa_{\rm cr} {\nabla}e_{\rm cr}
 \end{bmatrix}^\intercal \,,\quad
 \cS = \begin{bmatrix} 
  0\\
 \\
0\\
 \\
0
 \\
 \\
  -p_{\rm cr}{\nabla\cdot {\bf v}} \\
\end{bmatrix}
\end{eqnarray}
where
\begin{equation}\label{eq:eT}
  e_{\rm t} = e_{\rm g}  + e_{\rm cr}
            = \left(\frac{1}{2}\rho |{\bf v}|^2 + \frac{p_{\rm g}}{\gamma_{\rm g} - 1}\right)
              + \frac{p_{\rm cr}}{\gamma_{\rm cr} - 1} \,,
\end{equation}
and
\begin{equation}\label{eq:pT}
 p_{\rm t}  = p_{\rm g} + p_{\rm cr} 
\end{equation}
are, respectively, the total energy density (the sum of kinetic energy density, gas thermal energy density, and CR energy density $e_{\rm cr}$) and the total pressure (the sum of gas pressure $p_{\rm g}$ and CR pressure $p_{\rm cr}$). 
The adiabatic constants of the thermal and CR fluids are defined as $\gamma_{\rm g}=1+ p_{\rm g}/e_{\rm th}$ (where thermal energy density $e_{\rm th} = e_{\rm g} - \rho\,|{\bf v}|^2/2$) and $\gamma_{\rm cr}=1+ p_{\rm cr}/e_{\rm cr}$, and their values are taken to be $5/3$ and $4/3$ respectively.
Note that the last element of $\cU$ ($e_{\rm cr}$) does not obey a conservative equation. CRs can lose/gain energy due to the term $-p_{\rm cr}{\nabla\cdot{\bf v}}$ representing coupling between thermal and CR fluids. This term, involving a derivative, gives a non-zero and non-unique contribution across a shock which depends on numerical implementation. Later we explore various implementations of this term and the associated numerical challenges.
\begin{figure}
\centering
\includegraphics[height=1.8in,width=2.7in]{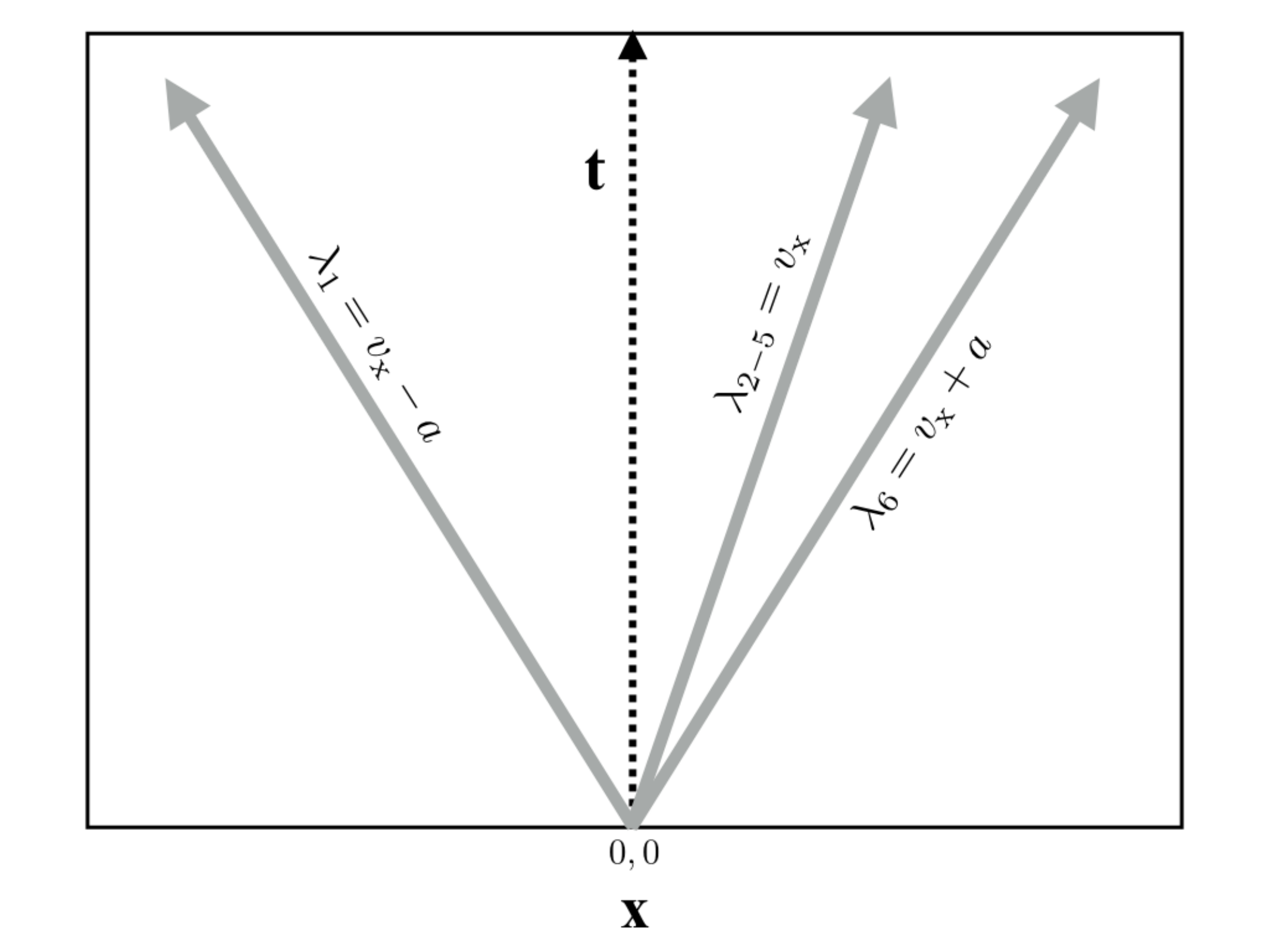}
\caption{Wave diagram of two-fluid (gas + CR) hydro system. Three straight lines originating from $(0,0)$ represent eigenvectors, where the eigenvalues are labeled by $\lambda_{\rm p}$. For details, see texts associated with Eqs (\ref{eq:eig}) and (\ref{eq:reigen}).} 
\label{fig:wave}
\end{figure}
\subsection{Characteristic structure}
In order to understand the characteristic structure of the two-fluid equations, we neglect the CR diffusion term (i.e., the term $\kappa_{\rm cr} {\nabla}e_{\rm cr}$ of Eq. \ref{eq:def})\footnote{Diffusion term is usually separately implemented using the standard parabolic schemes (e.g., \citealt{Vaidya2017}).} and focus only on the hyperbolic structure of the equations.
Considering 1D Cartesian coordinate with three velocity components, the primitive form of Eq. (\ref{eq:gen}) then becomes
\begin{equation} \label{eq:primeq}
  \pd{\cV}{t} + \tens{A}_{\rm p}(\cV) \pd{\cV}{x} = 0\ ,
\end{equation}
where 
\begin{equation}
\cV =\begin{bmatrix} 
\rho
\\
v_{\rm x}
\\
v_{\rm y}
\\
v_{\rm z}
\\
p_{\rm g}
\\
p_{\rm cr}
\end{bmatrix},\, 
\tens{A}_{\rm p} = \begin{bmatrix} 
  v_{\rm x} & \rho & 0 & 0 & 0 & 0 
  \\
0 & v_{\rm x} & 0 & 0 & \frac{1}{\rho} & \frac{1}{\rho}
\\
0 & 0 & v_{\rm x} & 0 & 0 & 0 
\\
0 & 0 & 0 & v_{\rm x} & 0 & 0
\\
0 & \rho a^2_{\rm g} & 0 & 0 & v_{\rm x} & 0
\\
0 & \rho a^2_{\rm cr} & 0 &0 & 0 & v_{\rm x}
\end{bmatrix}\ .
\end{equation}
Here $a_{\rm g}=(\gamma_{\rm g} p_{\rm g}/\rho)^{\HALF}$ and $a_{\rm cr}=(\gamma_{\rm cr} p_{\rm cr}/\rho)^{\HALF}$. Defining the effective sound speed as $a=(a^2_{\rm g} + a^2_{\rm cr})^{\HALF}$, the eigenvalues of the characteristic matrix $A_{\rm p}$ are found to be (also see Fig. \ref{fig:wave})
\begin{eqnarray}\label{eq:eig}
\lambda_{\rm p}= v_{\rm x}-a,\, v_{\rm x}, \, v_{\rm x},   \,v_{\rm x},\, v_{\rm x}, \,v_{\rm x}+a.
\end{eqnarray}
Right eigenvectors of the characteristic matrix $\tens{A}_{\rm p}$ can be written as,
\begin{eqnarray} \label{eq:reigen}
  \tens{R}_{\rm p}=\{r^{mn}_{\rm p}\} &=&\begin{bmatrix}
  1 & 1 & 0 & 0 & 0 & 1\\
  -\frac{a}{\rho} & 0 & 0 & 0 & 0 & \frac{a}{\rho}\\
  0 & 0 & 1 & 0 & 0 & 0 \\
  0 & 0 & 0 & 1 & 0 & 0\\
  a^2_{\rm g} & 0 & 0 & 0 & -1 & a^2_{\rm g}\\
  a^2_{\rm cr} & 0 & 0 & 0 & 1 & a^2_{\rm cr}
  \end{bmatrix} .
\end{eqnarray}

In the absence of CR fluid, the $5^{\rm th}$ column and the $6^{\rm th}$ row of the matrix $\tens{R}_{\rm p}$ are absent, i.e., the right-going eigenvalue is $\lambda=v_{\rm x}+a_{\rm g}$ and the corresponding eigenvector $r_{\rm 5}=\{1, a_{\rm g}/\rho, 0, 0, a_{\rm g}^{2}\}^\intercal$. This implies that the ratios between the fluctuations in density, velocity, and pressure scale as $1: a_{\rm g}/\rho: 0: 0: a_{\rm g}^{2}$. Therefore, any change in the x-velocity in a right going sound wave causes a simultaneous change in the density and the pressure, which increase/decrease without any ambiguity. However, the presence of CRs can introduce ambiguity in the fluctuations, which is illustrated as follows.

As shown in Eq. (\ref{eq:reigen}), for the right-going sound waves in the two-fluid CR equations the ratios between the fluctuations in density, velocity, thermal pressure, and CR pressure scale as $1: a/\rho: 0: 0: a_{\rm g}^{2}:a_{\rm cr}^{2}$, i.e., the fluctuation of the x-velocity can be satisfied for multiple values of $a_{\rm g}$ and $a_{\rm cr}$ [because $a=(a^2_{\rm g} + a^2_{\rm cr})^{\HALF}$].
This ambiguity affects the fluctuations of thermal and CR pressures. Moreover, the $5^{\rm th}$ column of the right eigenvector shows that at the contact discontinuity (represented by $r_{2}$ to $r_{5}$), the ratio between the fluctuations of the thermal pressure and CR pressure is $-1$. This means that, although the total pressure ($p_{\rm t}$, see Eq. \ref{eq:pT}) on both side of the contact discontinuity is constant, the thermal and CR pressures may not remain the same across a contact discontinuity. The fluctuation in thermal pressure across the contact discontinuity is compensated by a gain/loss of the CR pressure or vice versa. This subtlety in the characteristic structure of the two fluid CR equations leads to numerical issues addressed in the next sections.

\section{Non-uniqueness problem at shocks} \label{subsec:closure}
Related to the ambiguities discussed in the previous section, there are serious problems regarding the uniqueness of the solutions of the two-fluid equations at shocks. In two-fluid CR-HD, CR diffusion coefficient ($\kappa_{\rm cr}$) and CR adiabatic index ($\gamma_{\rm cr}$) need to be specified. These are generally known as the closure parameters, as they must be prescribed to close the system of equations (Eq \ref{eq:gen}). Although these choices are crucial to quantify the overall impact of CRs, their role is significantly important at shocks. 
For example, if one considers the time-evolution of CR distribution function (see e.g., \citealt{Duffy1994}), then a self-consistent treatment for $\gamma_{\rm cr}$ and $\kappa_{\rm cr}$ in the two-fluid model is necessary. However, the fluid equations are not sufficient to estimate these parameters (e.g., \citealt{Achterberg1984}). Here, we show that an additional equation of state is required to obtain a unique post-shock solution, even when a time-independent $\gamma_{\rm cr}$ and $\kappa_{\rm cr}$ are used (as is common in the two-fluid treatment). This EOS can be regarded as a closure across shocks when the length scale of interest is much larger than the shock transition scale. Since the galactic scales are orders of magnitude larger than the CR Larmor radius, below we neglect CR diffusion and focus on the far- upstream and downstream of a shock. In section \ref{sec:discussions}, we further discuss the role of cosmic ray diffusion in determining the small-scale structure of a two-fluid shock.

In the shock rest frame, the mass, momentum, and energy conservation equations for the two-fluid CR-HD model obey the Rankine-Hugoniot jump conditions
\begin{eqnarray}
\label{eq:mass_c}
\left[\rho v \right] & = & 0\ ,\\
\label{eq:mom_c}
\left[\rho v^2 + p_{\rm g} + p_{\rm cr}\right] & = & 0\ ,\, \\
\label{eq:engy_c}
\left[\left(\frac{1}{2}\rho v^2 + \frac{\gamma_{\rm g}}{\gamma_{\rm g}-1} p_{\rm g} + \frac{\gamma_{\rm cr}}{\gamma_{\rm cr}-1} p_{\rm cr} \right)\, v  \right]  &= &0
\end{eqnarray}
respectively, where $[\mathcal{A}]=\mathcal{A}_{\rm 1}-\mathcal{A}_{\rm 2}$ and the subscripts $1$ and $2$ represent the far-upstream and downstream fluid variables. 
These equations show that there are three conservation laws (for mass, momentum, and total energy) but four unknowns ($\rho$, ${v}$, $p_{\rm g}$, $p_{\rm cr}$). This constitutes {\it a closure problem} across shocks.

In order to highlight this problem, we introduce the compression ratio $\mathcal{R} = \rho_{\rm 2}/\rho_{\rm 1} = v_{\rm 1}/v_{\rm 2} $, and the upstream gas and CR Mach numbers, are respectively defined as
\begin{eqnarray}
\mathcal{M}_{\rm g, 1}  = \frac{v_{\rm 1}}{(\gamma_{\rm g}p_{\rm g, 1}/\rho_{\rm 1})^{\HALF}}\,, \quad  \mathcal{M}_{\rm cr, 1}  = \frac{v_{\rm 1}}{(\gamma_{\rm cr}p_{\rm cr, 1}/\rho_{\rm 1})^{\HALF}}\,,
\end{eqnarray} 
while the effective upstream Mach number of the composite fluid is $\mathcal{M}_{\rm 1} = (\mathcal{M}_{\rm g, 1}^{-2}+\mathcal{M}_{\rm cr, 1}^{-2} )^{-\HALF}$. We then normalize the gas and CR pressures relative to the upstream {\it gas pressure} and denote them by $\mathcal{P}_{\rm g, i} = p_{\rm g, i}/p_{\rm g,1}$ and $\mathcal{P}_{\rm cr, i} = p_{\rm cr, i}/p_{\rm g,1}$ respectively, where $i\in 1, 2$. 
The shock jump conditions can then be rewritten as
\begin{equation} \label{eq:mom_c2}
\begin{split}
\mathcal{M}_{\rm 1}^2 + \frac{1+ \mathcal{P}_{\rm cr,1}}{\gamma_{\rm g} +\gamma_{\rm cr}\mathcal{P}_{\rm cr,1}} =  \frac{\mathcal{M}_{\rm 1}^2}{\mathcal{R}} + \frac{\mathcal{P}_{\rm g,2} + \mathcal{P}_{\rm cr,2}}{\gamma_{\rm g} +\gamma_{\rm cr}\mathcal{P}_{\rm cr,1}}  
\end{split}
\end{equation}
\begin{equation}\label{eq:engy_c2}
\begin{split}
\mathcal{R}\left\{\frac{\mathcal{M}_{\rm 1}^2}{2} + \frac{1}{\gamma_{\rm g} +\gamma_{\rm cr}\mathcal{P}_{\rm cr,1}}\left( \frac{\gamma_{\rm g}}{\gamma_{\rm g}-1} + \frac{\gamma_{\rm cr}}{\gamma_{\rm cr}-1} \mathcal{P}_{\rm cr,1}\right)\right\} =\\ \left\{\frac{\mathcal{M}_{\rm 1}^2}{2\, \mathcal{R}} + \frac{1}{\gamma_{\rm g} +\gamma_{\rm cr}\mathcal{P}_{\rm cr,1}}\left( \frac{\gamma_{\rm g}}{\gamma_{\rm g}-1} \mathcal{P}_{\rm g,2} + \frac{\gamma_{\rm cr}}{\gamma_{\rm cr}-1} \mathcal{P}_{\rm cr,2}\right)\right\} ,
\end{split} 
\end{equation}
where the normalized upstream CR pressure, $\mathcal{P}_{\rm cr,1}$ is \begin{eqnarray}\label{eq:pcr1_nom}
\mathcal{P}_{\rm cr,1} = \frac{p_{\rm cr, 1}}{p_{\rm g, 1}} =\left(\frac{\gamma_{\rm g} }{ \gamma_{\rm cr}}\right) \left(\frac{\mathcal{M}_{\rm g, 1}}{\mathcal{M}_{\rm cr, 1}}\right)^2 \ .
\end{eqnarray}
Eqs (\ref{eq:mom_c2}) and (\ref{eq:engy_c2}) must be solved for the downstream state ($\mathcal{P}_{\rm g,2}$, $\mathcal{P}_{\rm cr,2}$, and $\mathcal{R}$), given the upstream conditions. This is an underdetermined system since one has more unknowns than equations.

To get a unique solution, we need an additional equation and this is the closure problem. We can fix the downstream CR pressure ($\mathcal{P}_{\rm cr,2}$) in several ways. Here we briefly discuss three possible equation-of-states (EoSs; see Appendix \ref{app:eos_detail} for details). 

\begin{itemize}
\item $w_{\rm cr}$-EoS: 
In this case, the downstream CR pressure is set to a fraction $w_{\rm cr}$ of the downstream total pressure, i.e.,
\begin{eqnarray} \label{eq:wcr}
w_{\rm cr} = \frac{p_{\rm cr, 2}}{p_{\rm g, 2}+p_{\rm cr, 2}}= \frac{\mathcal{P}_{\rm cr,2}}{(\mathcal{P}_{\rm g,2}+ \mathcal{P}_{\rm cr,2})}\, ,
\end{eqnarray}
where $0\leq w_{\rm cr}\leq 1$. 
The post-shock solution depends on the  prescribed value of $w_{\rm cr}$ (Appendix \ref{app:wcr}).
\item $\epsilon_{\rm cr}$-EoS:
%
For this EoS, the downstream CR enthalpy flux is set as a fraction $\epsilon_{\rm cr}$ of the upstream total energy flux, i.e.,
\begin{eqnarray} \label{eq:pcr1}
 \epsilon_{\rm cr} & = & \frac{\frac{\gamma_{\rm cr}}{\gamma_{\rm cr}-1} p_{\rm cr, 2} v_{\rm 2} }{ \left(\frac{1}{2}\rho_{\rm 1}v_{\rm 1}^2 + \frac{\gamma_{\rm g}}{\gamma_{\rm g}-1} p_{\rm g,1} + \frac{\gamma_{\rm cr}}{\gamma_{\rm cr}-1} p_{\rm cr,1} \right)\, v_{\rm 1}},
\end{eqnarray}
where $0\leq\epsilon_{\rm cr}\leq 1$ ensures energy conservation. In Appendix \ref{app:rela_wcrecr}, we show that w$_{\rm cr}$-EOS and $\epsilon_{\rm cr}$-EOS are interchangeable and therefore the shock solutions are expected to be similar in these two cases.
\item Adiabatic-EoS:
%
In this case, the CR entropy across a shock is assumed to be constant, i.e., the downstream CR pressure can be set to
\begin{eqnarray} \label{eq:eos-adia}
p_{\rm cr, 2} = p_{\rm cr, 1} \left(\frac{\rho_{\rm 2}}{\rho_{\rm 1}}\right)^{\gamma_{\rm cr}} \rightarrow \mathcal{P}_{\rm cr, 2} = \mathcal{P}_{\rm cr, 1} \mathcal{R}^{\gamma_{\rm cr}}.
\end{eqnarray}
A method for obtaining post-shock solution for this EoS is discussed in Appendix \ref{app:adia}. This EoS is sometimes used to check consistency between analytic solution and test problem in numerical simulation (see e.g., \citealt{Pfrommer2006}; \citealt{Kudoh2016}). However, as we show here, this is only one of the possible EoSs. 
\end{itemize}

Despite the closure problem discussed here, the CR two-fluid equations are mostly used without an EoS imposed across shocks. The impact of the closure problem in numerical implementation, as well as in the resulting solutions, are presented in the next sections.

\BLACK

\begin{table*}
\renewcommand{\arraystretch}{1.2}
\setlength{\arrayrulewidth}{0.15mm}
\setlength{\doublerulesep}{0.2mm}
\caption{
Various methods for implementing the non-conservative term of CR-HD equations$^{\dagger}$.}
\label{fig:numericalimp}
\begin{footnotesize}
\centering
\begin{tabular}{|c|c|c|c|}
\hline
\multicolumn{2}{|c|}{\textbf{Assumptions used}}  &\textbf{Coupling terms}  &  \textbf{Nomenclature \& References e.g.} \\
\multicolumn{2}{|c|}{\textbf{in numerical implementation}} & &     \\
\hline
\multirow{8}{*}{} 
& & {\bf A.\,\,p\,dv} & {\bf A1. \emph{Eg+Ecr (OpSplit-pdv)}}\\
& & & \citealt{Pfrommer2006,Sharma2009};\\
& & $+p_{\rm cr}\nabla\cdot {\bf v}$& \citealt{Salem2014,Butsky2018}\\ 
\cline{4-4}
& {\bf Option 1:  $Eg + Ecr$} & $-p_{\rm cr}\nabla\cdot {\bf v}$ & {\bf A2. \emph{Eg+Ecr (Unsplit-pdv)}}\\
&  Gas energy + CR energy eqs  &  & This work\\
\cline{3-4}
& & {\bf B.\,\,v\,dp} &  {\bf B1. \emph{Eg+Ecr (OpSplit-vdp)}}\\
& & $-{\bf v}\cdot\nabla p_{\rm cr}$ &\\
\cline{4-4}
& & $+{\bf v}\cdot\nabla p_{\rm cr}$ & {\bf B2. \emph{Eg+Ecr (Unsplit-vdp)}}\\
& & &\\
\cline{2-4}
\multirow{8}{*}{\rotatebox{90}{\hspace{5in} {\Large Two-fluid CR-HD} \hspace{-0.7in}}}
& & {\bf A.\,\,p\,dv} & {\bf A1. \emph{Et+Ecr (OpSplit-pdv)}}\\
& & $0$& \citealt{Yang2012}; \\ 
\cline{4-4}
& {\bf Option 2:  $Et + Ecr$} & $-p_{\rm cr}\nabla\cdot {\bf v}$ & {\bf A2. \emph{Et+Ecr (Unsplit-pdv)}}\\
& Total energy + CR energy eqs &  & This work\\
\cline{3-4}
& & {\bf B.\,\,v\,dp} &  {\bf B1. \emph{Et+Ecr (OpSplit-vdp)}}\\
& & $0$ & \\
\cline{4-4}
& & $+{\bf v}\cdot\nabla p_{\rm cr}$ & {\bf B2. \emph{Et+Ecr (Unsplit-vdp)}}\\
& & &\\
\cline{2-4}
\multirow{4}{*}{} 
& {\bf  Option \,3:  $Et + Scr$} &  & {\bf  \emph{Et+Scr}}\\
& Total energy + CR  entropy eqs  & &  \citet{Kudoh2016}\\ 
\hline
\end{tabular}
\end{footnotesize}
\\
\vspace{1em}
{\raggedright $^{\dagger}$Note: Three options differ w.r.t. to the implementation of gas/total energy and CR energy equations. Option $1$: \emph{Eg + Ecr}, the energy density of CRs [$e_{\rm cr}=p_{\rm cr}/(\gamma_{\rm cr}-1)$] and gas [$e_{g} = \rho\,|{\bf v}|^2/2 + p_{\rm g}/(\gamma_{\rm g}-1)$] are separately evolved, where the coupling term is taken either as $p_{\rm cr} {\nabla \cdot \bf{v}}$ ($\emph{pdv}$ method) or ${\bf v \cdot \nabla} p_{\rm cr}$ ($\emph{vdp}$ method). Option $2$: \emph{Et + Ecr}, the total energy density, i.e., $(e_{\rm g} + e_{\rm cr})$ is used instead of $e_{\rm g}$ in order to update the energy density of the system and the coupling term can either be $\emph{pdv}$ or $\emph{vdp}$, similar to Option $1$. However, in Option $2$, the coupling term does not appear in the total energy equation. Option $3$: \emph{Et + Scr}, CR energy density is updated from the advection of a passive scalar and coupling terms are absent. In options $1$ and $2$, depending on the implementation of the coupling terms, they can be divided into two sub methods: (i) operator-splitting (\emph{OpSplit}) and (ii) \emph{Unsplit}. Naming of different methods is given in the rightmost column of this table.
\par}
\end{table*}
\section{Numerical framework} \label{sec:method}
In this section, we explore a variety of different discretization strategies for the numerical solution of the two-fluid equations. All methods differ essentially in the representation of the coupling term (the non-conservative term in Eq. \ref{eq:def}).
Let $\cL_h()$ and $\cL_s()$ be the discrete operators corresponding to the evolution of the homogeneous part of the equations and to the source term alone, respectively. The update step can then be achieved via operator splitting,
\begin{equation}\label{eq:OpSplit}
  \begin{array}{l}
  \DS \quad \cU^{*} = {\cal L}_h(\cU^n)   \\ \noalign{\medskip}
  \DS \quad \cU^{n+1} = {\cal L}_s(\cU^*) \,,
  \end{array}
\end{equation}
or in a fully unsplit fashion:
\begin{equation}\label{eq:Unsplit}
  \cU^{n+1} = ({\cal L}_h+{\cal L}_s)(\cU^n)  \,.
\end{equation}
We label these two approaches as \emph{OpSplit} (Eq. \ref{eq:OpSplit}) and \emph{Unsplit} (Eq. \ref{eq:Unsplit}), respectively. 
Notice that the operator split method presented here is only first-order accurate in time, but it can be made formally second order accurate using Strang (or alternate) splitting (\citealt{Strang1968}).

\begin{figure}
\centering
\includegraphics[height= 1.9in,width=3.5in]{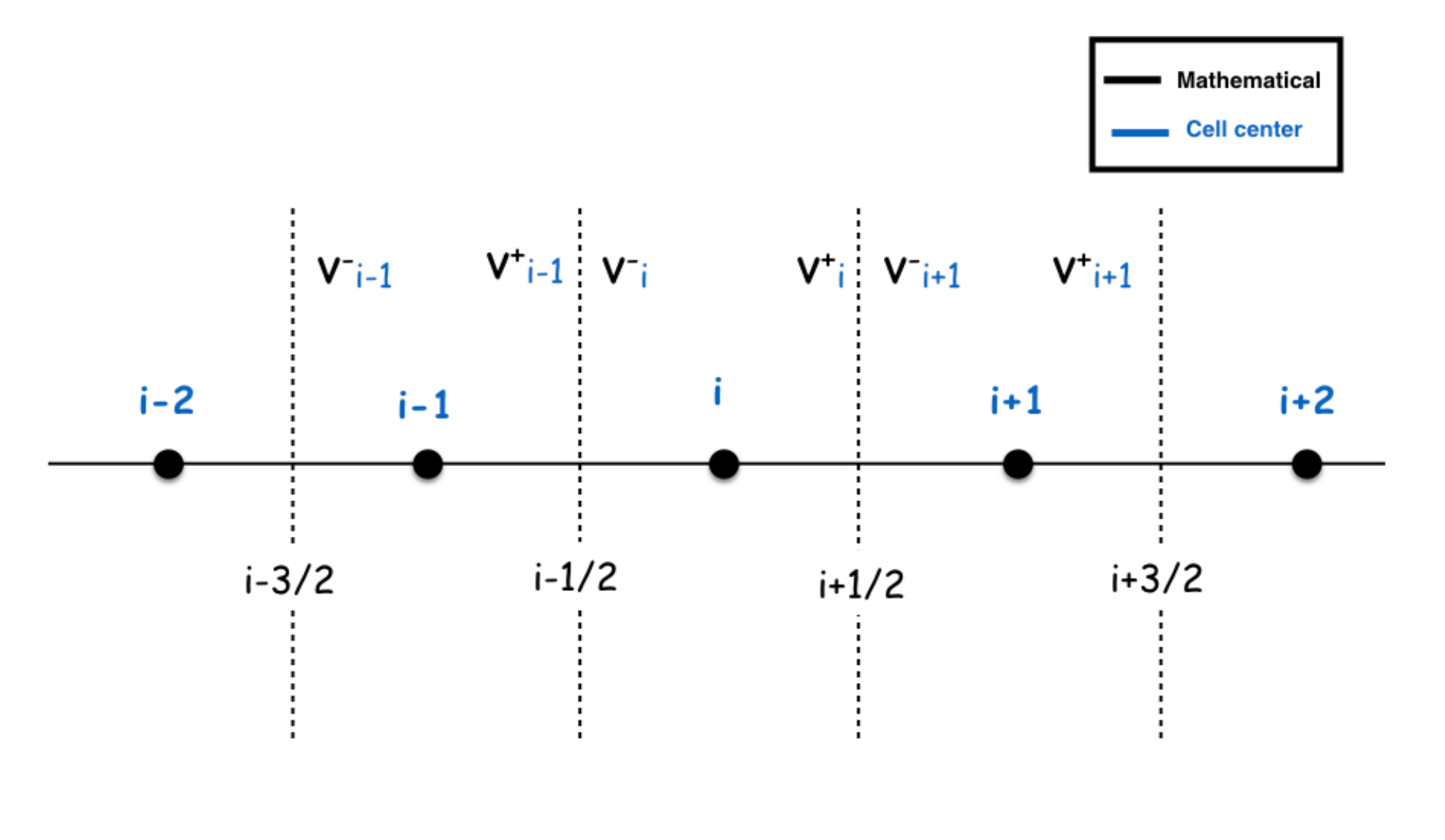}
\caption{Schematic representation of the grid. Black circles denote zone-centers (labeled with integer notation, $i$) , while dashed vertical lines represent cell interfaces (labeled with half-integer notation, $i+\HALF$). $V^{-}$ and $V^{+}$ represent, respectively, the reconstructed value of a fluid quantity $V$ to the left ($-$) or right ($+$) with respect to the zone center. For $1^{\rm st}$-order reconstruction, $V^-_i = V^+_i = V_i$. 
}
\label{fig:nom}
\end{figure}

The rationale for choosing $\cL_h$ and $\cL_s$ depends on the implementation of fluid energy densities and for this reason we classify them into three different options: Option $1$ - \quotes{Eg+Ecr}, $2$ - \quotes{Et+Ecr}, and $3$ - \quotes{Et+Scr}. The differences between various methods are briefly illustrated in Table \ref{fig:numericalimp} and the nomenclature of different indices is shown in Fig. \ref{fig:nom}. Note that since the implementation of mass and momentum equations remain the same in all of the methods, we focus solely on the energy equations.

\subsection{Option 1: Eg+Ecr} \label{subsec:method1}
Here the energies of thermal and CR fluids, coupled by the source term, are evolved separately; i.e., we solve the following equations
\begin{eqnarray}\label{eq:apdv1}
\noindent \pd{e_{\rm g}}{t} & = &
  - {\nabla} \cdot\left[(e_{\rm g} + p_{\rm t}){\bf v}\right] + p_{\rm cr} {\nabla} \cdot {\bf v} ,
  \\ \label{eq:apdv2}
  \pd{e_{\rm cr}}{t}  & = &   - {\nabla} \cdot(e_{\rm cr}{\bf v}) - p_{\rm cr} {\nabla}\cdot {\bf v} \,.
\end{eqnarray}
By suitably rewriting the RHS of Eqs. (\ref{eq:apdv1}) and (\ref{eq:apdv2}), the coupling term may be implemented either as $p_{\rm cr} {\nabla \cdot \bf{v}}$ (\emph{pdv} method) or ${\bf v \cdot \nabla} p_{\rm cr}$ (\emph{vdp} method). Although the former choice (\emph{pdv} method) is found to be numerically robust and more common, we discuss both implementations separately in the following sections. Note that total energy conservation is ensured by adopting the same discretization for the source term in Eq. (\ref{eq:apdv1}) and (\ref{eq:apdv2}).
\begin{center}
{\bf A. $\emph{pdv}$ method}
\end{center}
In this method, the CR energy flux is defined, following Eq. (\ref{eq:apdv2}) by ${\bf F}_{\rm cr} = e_{\rm cr}{\bf v}$. Since the source term is $p_{\rm cr} {\nabla} \cdot {\bf v}$, we refer to this method as the \emph{pdv} method. 

\noindent{\bf A1. \emph{OpSplit-pdv} method:}
In many studies, the $p_{\rm cr} {\nabla \cdot \bf{v}}$ term has been implemented using an operator splitting method (e.g., \citealt{Pfrommer2006}; \citealt{Sharma2009}; \citealt{Salem2014}; \citealt{Gupta2018a}). 
Following Eq. (\ref{eq:OpSplit}), $e_{\rm cr}$ is first evolved along with the other hydrodynamic variables to yield $e^*_{\rm cr}$. This value is then used to calculate $p_{\rm cr} {\nabla} \cdot {\bf v}$ at the next step. 
For a $1^{\rm st}$-order time-stepping scheme this amounts to
\begin{eqnarray}
  \label{eq:op1pdv}
  e^{*}_{{\rm cr,} i}& = & e^{n}_{{\rm cr,} i} - \Delta t\left(
   \frac{f^{n}_{i+\HALF} - f^{n}_{ i-\HALF}}
                                {\Delta x_i}\right) \,, \\
   \label{eq:op2pdv}
e^{n+1}_{{\rm cr,} i} & = & e^{*}_{{\rm cr,} i} - \Delta t\,
                      p^{*}_{{\rm cr,} i}\,\left(
                   \frac{v^{*}_{i+\HALF}-v^{*}_{i-\HALF}}
                        {\Delta x_i}\right) \,,
\end{eqnarray}
where, in a Godunov-type finite volume scheme, $f^{n}_{i+\HALF}$ represents an upwind flux computed with a Riemann solver. Here we employ the HLL Riemann solver (see Chapter $10$ in \citealt{Toro2009}) to evaluate the fluxes of the conservative terms, while the interface velocity (see Fig. \ref{fig:numericalimp}) used for the non-conservative term in Eq. (\ref{eq:op2pdv}) is defined as 
\begin{equation} \label{eq:vop}
   v^{*}_{i+\HALF} = \frac{1}{2}(v^{*}_{i}+v^{*}_{i+1}) \,.
\end{equation}
In other words, the quantity ${\nabla} \cdot {\bf v}$ is estimated using a cell centered method
\footnote{Note that at the time of updating Eqs. (\ref{eq:apdv1}) and (\ref{eq:apdv2}) without $p_{\rm cr} {\nabla \cdot \bf{v}}$, the Riemann solver does not have any information about the term $p_{\rm cr} {\nabla} \cdot {\bf v}$.
We find that such splitting modifies the effective sound speed to $a= [(a^2_{\rm g} + (\gamma_{\rm g}/\gamma_{\rm cr})a^2_{\rm cr}]^{\HALF}$, instead of $a= [a^2_{\rm g} + a^2_{\rm cr}]^{\HALF}$ (see Appendix \ref{app:A} for the derivation). 
The effective sound speed is required for computing the signal speed/time step of the Riemann solver.}. 
The gas energy equation (\ref{eq:apdv1}) is updated using a similar discretization as Eqs. (\ref{eq:op1pdv}) and (\ref{eq:op2pdv}). Eq. (\ref{eq:op1pdv}) is readily generalized to second-order temporal acuracy by using a Runge-Kutta method already available in the code while direct extension to curvilinear geometries is thoroughly described in the work by \citet{Mignone.2014}.

\noindent{\bf A2. \emph{Unsplit-pdv} method:}
In the fully unsplit scheme (Eq. \ref{eq:Unsplit}), the $p_{\rm cr} {\nabla \cdot \bf{v}}$ term is directly added to the RHS of Eqs. (\ref{eq:apdv1}) and (\ref{eq:apdv2}). This yields, for the CR energy equation, the following update:
\begin{equation} \label{eq:ecrup}
e^{n+1}_{{\rm cr,} i} = e^{n}_{{\rm cr,} i}
    - \Delta t\left[\left(\frac{f^{n}_{i+\HALF}-f^{n}_{i-\HALF}}
                         {\Delta x_i}\right)
   +  \, p^{n}_{{\rm cr,} i}\,
     \left(\frac{v^{n}_{i+\HALF} - v^{n}_{i-\HALF}}
          {\Delta x_i}\right)\right] \,.
\end{equation}
Although several options are possible for choosing $p^{n}_{{\rm cr,} i}$ and $v^{n}_{i+\HALF}$ (see e.g., Appendix \ref{app:B}), one can take advantage of the Riemann solver information to estimate  $p^{n}_{{\rm cr,} i}$ and $v^{n}_{\rm i+\HALF}$. 
We find that a robust choice for the cell interface velocity is
\begin{equation} \label{eq:v}
  v^{n}_{\rm i+\HALF}= \frac{\cU^{n}_{[m_x],  i + \HALF}}
                            {\cU^{n}_{[\rho], i + \HALF}} \,,
\end{equation}
where $ \cU_{[m_x]} $ and $\cU_{[\rho]}$ are the momentum- and density- state variables obtained from the HLL Riemann solver (\citealt{Toro2009}). 

The CR pressure in Eq. (\ref{eq:ecrup}) can also be chosen in various ways. We find that {\BLUE a} robust selection is
\begin{equation} \label{eq:pcr}
p^{n}_{\rm cr, i} = \frac{\gamma_{\rm cr}-1}{2}
     \left(  e^{n}_{{\rm cr}, i-\HALF}
           + e^{n}_{{\rm cr}, i+\HALF}\right)\ ,
\end{equation} 
where $e^{n}_{\rm cr, i+\HALF}$ is the state-variable obtained from the  HLL Riemann solver, without including the $p_{\rm cr}  \nabla \cdot {\bf v}$ term. We label this method as ``\emph{Eg+Ecr (Unsplit-pdv)}". 

One of the advantages of choosing the HLL solver is that it does not distinguish between the left- and right- states of the contact wave. This helps avoid a selection ambiguity in the $5^{\rm th}$ eigenvector as discussed in section \ref{sec:basics}. Thus the choice of HLL Riemann solver is motivated by simplicity and consistency in treating the fluxes and the source terms.
\begin{center}
{\bf B. $\emph{vdp}$ method}
\end{center}
Alternatively, the coupling term can be implemented in ${\bf v \cdot \nabla}p_{\rm cr}$ form. 
The evolution of CR energy density is then obtained by rewriting the CR energy equation (Eq. \ref{eq:apdv2}) as
\begin{equation}\label{eq:bvdp}
 \pd{e_{\rm cr}}{t} = -{\nabla} \cdot\left[( e_{\rm cr} + p_{\rm cr}) {\bf v}\right]
                      +\BLACK{\bf v \cdot \nabla}p_{\rm cr} \,,
\end{equation}
which, at the discrete level, becomes similar to Eq. (\ref{eq:op1pdv}) and (\ref{eq:ecrup}) by exchanging $\bf v$ and $p_{\rm cr}$.

Although, Eqs. (\ref{eq:apdv2}) and (\ref{eq:bvdp}) are mathematically equivalent, the numerical discretizations are not the same. 
Since the coupling term now is ${\bf v \cdot \nabla}p_{\rm cr}$, we call this method as \emph{vdp} method. As before, this coupling term can be included via an operator splitting method (\emph{OpSplit-vdp}) or an unsplit method (\emph{Unsplit-vdp}).

One may now choose a cell-centered discretization for ${\bf v}_i$ and the arithmetic average between the left and right interface values for $p_{{\rm cr}, i+\HALF}$.
However, as it will be shown in section \ref{sec:result}, this choice leads to results that depend on the type and order of the spatial reconstruction algorithm. 
This effect is considerably less pronounced with the previous ($pdv$) method.
\subsection{Option 2: Et+Ecr}\label{subsec:method2}
In this case, we replace Eq. (\ref{eq:apdv1}) with the total energy equation
\begin{equation}\label{eq:pdv2a}
  \pd{e_{\rm t}}{t}= - {\nabla} \cdot\left[(e_{\rm t} + p_{\rm t}){\bf v}\right]
  \,, 
\end{equation}
where $e_{\rm t}$ and $p_{\rm t}$ are given by Eqs. (\ref{eq:eT}) and (\ref{eq:pT}).

The total energy density now directly obeys a conservative equation. However, the CR energy equation still contains the coupling term, which can be implemented using the \emph{pdv} or \emph{vdp} method as described earlier. After extensive numerical testing, we have found that the numerical results are consistent for various  reconstruction schemes only with the $pdv$ method in a fully unsplit fashion, and $v$ and $p_{\rm cr}$ are chosen as given in Eqs. (\ref{eq:v}) and (\ref{eq:pcr}).
We thus label this method as ``\emph{Et+Ecr (Unsplit-pdv)}". It is worth mentioning that both ``\emph{Et+Ecr (Unsplit-pdv)}" and ``\emph{Eg+Ecr (Unsplit-pdv)}" give an identical result when Eqs. (\ref{eq:v}) and (\ref{eq:pcr}) are used.

\begin{table*}
\centering
\caption{Test Problems}
\begin{tabular}{ llclc}
  \hline\hline
  Test & Initial Conditions    & $t_{\rm stop}$ & $N_x$ & Figure  \\
       & $\{\rho, v_{\rm x},p_{\rm g}, p_{\rm cr}\}$ &                &            & \\
\hline
 \GREEN $1$\BLACK. Pressure balance & $\begin{array}{ll}
                     {\rm L}: \; \{1  ,\, 1,\, 0.1,   \, 0.9\} \\ 
                     {\rm R}: \; \{1,\, 1,\, 0.9,\, 0.1\} 
                     \end{array}$ 
                   & $1.0$ & $200$ & \ref{fig:pbalance} \\
\hline 
  \GREEN $2$\BLACK. Shock tube A &  $\begin{array}{ll}
                     {\rm L}: \; \{1  ,\, 0,\, 2,   \, 1\}\\ 
                     {\rm R}: \; \{0.2,\, 0,\, 0.02,\, 0.1\} 
                   \end{array}$  
               & $0.1$ & $1000$ & \ref{fig:sodcr1_alldiff}, \ref{fig:sodcr1zoom}, \ref{fig:sodcr1} \\

\hline
 $3$. Shock tube B & $\begin{array}{ll}
                     {\rm L}: \; \{1  ,\, 0,\, 6.7\times 10^4,\, 1.3\times 10^5\} \\ 
                     {\rm R}: \; \{0.2,\, 0,\, 2.4\times 10^2,\, 2.4\times 10^2\} 
                  \end{array}$
              & $10^{-4}$ & $1000$  & \ref{fig:sodcr2} \\                                    
  \hline
  $4$. Blast wave & see section \ref{subsec:multiD}     & section \ref{subsec:multiD} & section \ref{subsec:multiD} & \ref{fig:blastcr} \\                                                                
  \hline
\end{tabular}\\
\raggedright{\footnotesize{Note: For all problems (except for ``Pressure balance" test problem where boundaries are periodic), boundary conditions are set to outflow.
Problems $1$-$3$ are performed in $1$D Cartesian geometry while for test $4$ we employ $1$D spherical and $3$D Cartesian geometries.}}
\label{tab:testproblems}
\end{table*} 
\subsection{Option 3: Et+Scr} \label{subsec:method3}
\citet{Kudoh2016} suggested that the difficulties related to the presence of the coupling term may be avoided if the evolution of CR energy density is obtained from the advection of a passive scalar. They redefined the CR pressure as $p_{\rm cr}\equiv \rho_{\rm cr}^{\gamma_{\rm cr}}$ and used a passive scalar: $\chi=\rho_{\rm cr}/\rho\equiv p_{\rm cr}^{1/\gamma_{\rm cr}}/\rho$ to update the CR energy equation.
Under this formalism, Eq. (\ref{eq:apdv2}) or (\ref{eq:bvdp}) reads
\begin{eqnarray} \label{eq:crden}
\frac{\partial}{\partial t}\left(\chi \rho\right) +{\nabla} \cdot\left(\chi \rho\,{\bf v}\right) = 0 \ .
\end{eqnarray}
The remaining equations maintain the same form as for {\it Option $2$}.
The advantage of this method is that the two-fluid equations become manifestly conservative, which makes the application of Godunov-type formalism straightforward.

Since Eq. (\ref{eq:crden}) is a tracer equation, the CR entropy $S_{\rm cr} \equiv p_{\rm cr}/\rho^{\gamma_{\rm cr}}$ does not experience a jump across a shock front\footnote{Entropy conservation also holds for the Euler equation in smooth regions but the conservative/integral form of the equations must be evolved at shocks for a physical solution.}. However, we argue that CRs, like the thermal plasma, should not be adiabatic across a shock. Indeed, CRs are accelerated with non-negligible efficiencies across strong shocks, but this possibility is not allowed by the strict isentropic evolution imposed by the above scalar equation. Isentropic evolution of CRs across shocks is discussed as one of the equation-of-states in section \ref{subsec:closure} (Eq. \ref{eq:eos-adia}). The results from this method are labeled by ``\emph{Et+Scr}''.


\subsection{Shock detection} \label{subsec:shockdetect}
%
In previous sections, we have discussed various possible numerical schemes for implementing CR energy equation. Since the coupling term in this equation involves a spatial derivative, both the numerical and physical  two-fluid shock jump condition can be non-unique. While a shock is generally defined as a surface where a sharp transition between upstream and downstream occurs, in numerical simulations, a shock is often broadened over several grid zones due to non-negligible numerical viscosity. To impose an EoS across a shock, it is important to discuss how one can detect shocked zones. For shock detection, we use the following three conditions:
\begin{enumerate}
\item Compressibility: $\nabla\cdot {\bf v} < 0$,
\item Pressure jump: $\left(\nabla p_{\rm t}\cdot \Delta {\bf x}\right)/ p_{\rm t, min} \geq \delta_{\rm threshold}$, and
\item Bypassing spurious waves at contact discontinuity which may fulfil conditions (i) and (ii): $\nabla T\cdot \nabla \rho > 0$,
\end{enumerate}
where $T$ is the fluid temperature  (see e.g., section $3.1.1$ in \citealt{Pfrommer2017}; also see Appendix B in \citealt{Mignone2012}). We find that the choice $\delta_{\rm threshold}=0.5-1$ works quite well for the test problems discussed in this work (see e.g., the top panel of Fig. \ref{fig:blastcr}).
%
\begin{figure*}
\centering
\includegraphics[height=3.9in,width=6.5in]{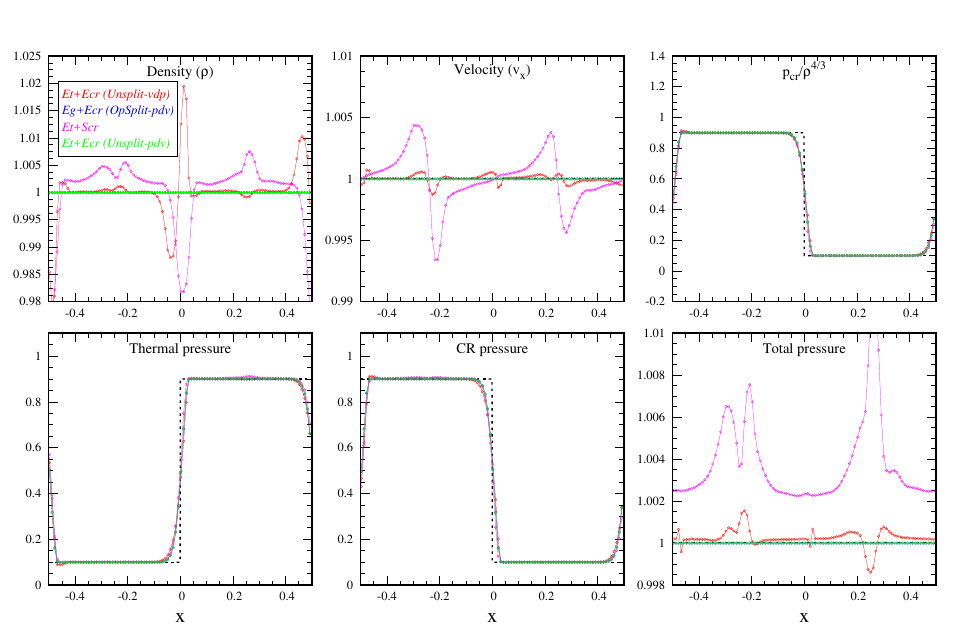}
\caption{Pressure balance test problem. Black dashed lines show the initial condition while circles represent various profiles after one box crossing at $t=1$. We employ linear reconstruction, RK2 time stepping and a CFL number of $0.6$ for all schemes. Equilibrium is maintained for $t>0$ only with {\emph{OpSplit-pdv}} (blue) and \emph{Unsplit-pdv} (green), while \emph{Unsplit-vdp} (red) and \emph{Et+Scr} (magenta) generates spurious waves. The vdp method is not preferred over pdv because although velocity is continuous across $x= 0$, CR pressure is not and therefore taking its derivative produces spurious disturbances at the contact discontinuity.}
\label{fig:pbalance}
\end{figure*}
%
\section{Test problems \& results}\label{sec:result}
In this section we compare the previously presented numerical methods for the solution of selected one- and multi-dimensional benchmarks. We briefly outline the used methods in the following:
\begin{itemize}
\item \emph{Eg+Ecr (OpSplit-pdv)} [Eqs \ref{eq:op1pdv} - \ref{eq:vop}]
\item \emph{Et+Ecr (Unsplit-vdp)}  [Eq. \ref{eq:bvdp}],
\item \emph{Et+Ecr (Unsplit-pdv)}  [Eqs. \ref{eq:ecrup} - \ref{eq:pcr}, and \ref{eq:pdv2a}], and
\item \emph{Et+Scr} (Eq. \ref{eq:crden}).
\end{itemize} 
As we shall show, the ``\emph{Et+Ecr (Unsplit-pdv)}'' method appears to be less sensitive to numerical details (spatial reconstruction, time-stepping, and CFL number). 
The initial conditions for the selected test problems are listed in Table \ref{tab:testproblems}. 
Computations are performed using either a $1^{\rm st}$-order scheme (Euler time stepping with flat reconstruction), a $2^{\rm nd}$-order scheme (RK2 time stepping with linear reconstruction) or a $3^{\rm rd}$-order scheme (RK3 time stepping with WENO reconstruction). The fiducial CFL number is set to $C_{\rm a} = 0.6$ unless otherwise stated.

\begin{figure*}
\centering
\includegraphics[height=4.2in,width=6.7in]{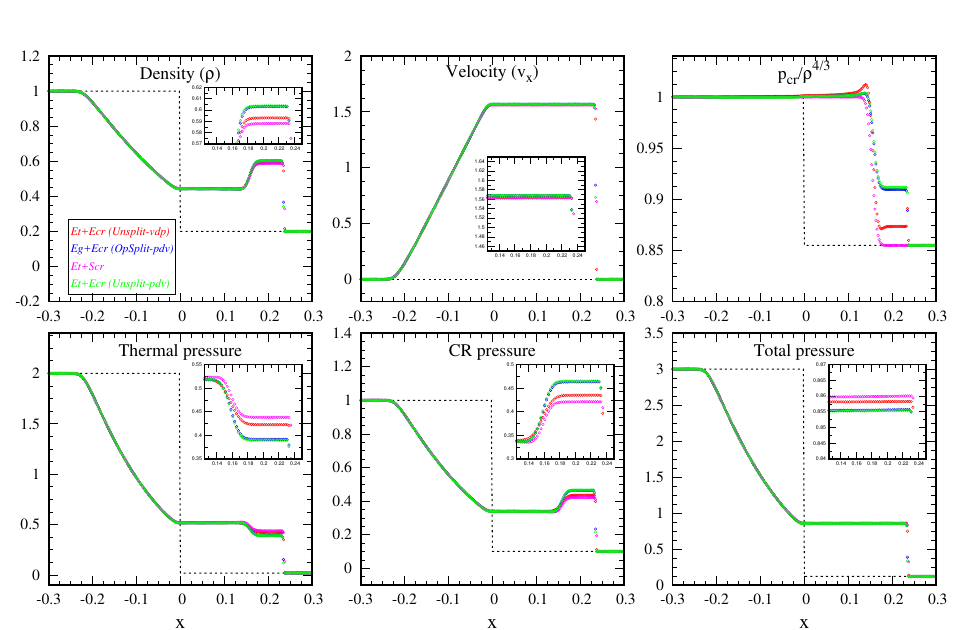}
\vspace{-1em}
\caption{Snapshots of fluid variables for the two-fluid shock tube A (see problem $2$ in Table \ref{tab:testproblems}). The black dashed curves represent the profiles at $t=0$, and while different colours correspond to the solution obtained at $t=0.1$ with the methods listed in the legend. Computations have been obtained using the $1^{\rm st}$-order scheme. The insets display the zoomed-in view of the post-shock region showing that different methods do not give a unique solution.}
\label{fig:sodcr1_alldiff}
\end{figure*}
\subsection{Pressure balance mode}\label{subsubsec:Pmod}
%
This is an important test problem designed to check the evolution of a pressure balance mode. 
At $t=0$, the left and right states are defined as $(\rho, v_{\rm x,} p_{\rm g}, p_{\rm cr})_{\rm L}=(1, 1, 0.1, 0.9)$ and $(\rho, v_{\rm x,}, p_{\rm g}, p_{\rm cr})_{\rm R}=(1, 1, 0.9, 0.1)$. 
Since, the total pressure is the same across the interface ($p_{\rm g}+p_{\rm cr}=1$), the profiles should not change in time. The pressure balance test is applicable to any multi-fluid system because of its trivial analytical solution. 

To obtain numerical solutions, we use the $2^{\rm nd}$-order scheme and impose periodic boundary conditions. The snapshots of fluid quantities at $t=1$ (i.e., after one advection time across the domain) are shown in Fig. \ref{fig:pbalance}. \emph{OpSplit-pdv} (blue) and  \emph{Unsplit-pdv} (green) curves do not show spurious oscillations. However, \emph{Unsplit-vdp} (red)\footnote{For the \emph{Unsplit-vdp} method, one can find a suitable combination of $p_{\rm cr}$ and $v$, which does not show these spurious oscillations. However, we find that the same choice does not provide a good solution for other test problems.} and \emph{Et+Scr} (magenta) schemes fail to maintain the pressure balance mode (for both HLL and HLLC Riemann solvers).

\begin{figure*}
\centering
\includegraphics[height=3.9in,width=6.7in]{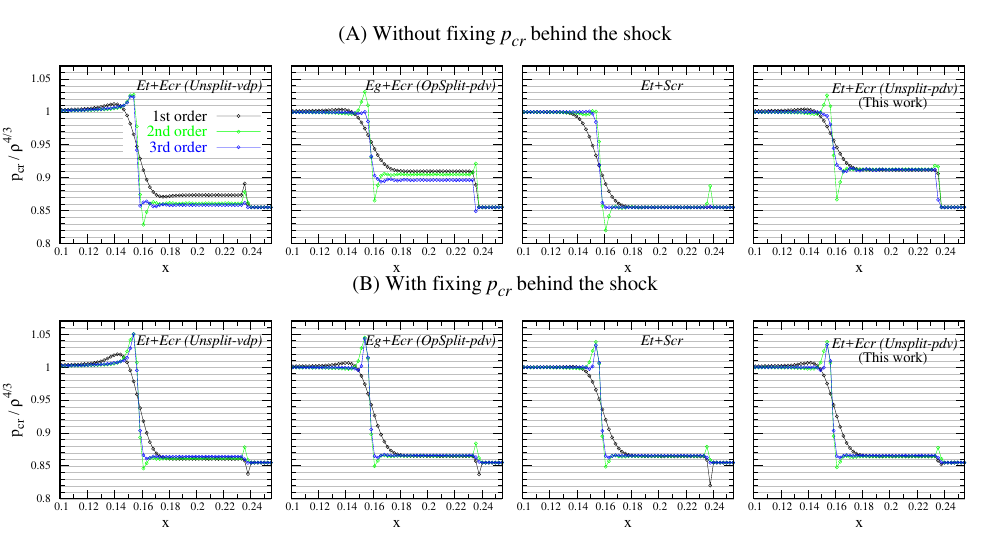}
\vspace{-1em}
\caption{The zoomed-in view of CR entropy ($p_{\rm cr}/\rho^{4/3}$) from shock tube A. Four vertically aligned panels stand for the different methods. For each of these methods, we show the results using the best possible combination of $p_{\rm cr}$ and $v$ from our experiments, except for the $3$rd column where the coupling term is not present. The three colours correspond to computations obtained with the $1^{\rm  st}-$ (black), $2^{\rm  nd}-$ (green), and $3^{\rm  rd}-$ (blue) order numerical schemes, respectively. In the top panel, the CR pressure is evolved using the equation. In the bottom panels, the post-shock CR pressure has been fixed using Eq. (\ref{eq:shk}) with $w_{\rm cr} = 0.5$. This result demonstrates that if the energy exchange between thermal and CR fluid at the shock is fixed, then all methods converge to the same solution.}
\label{fig:sodcr1zoom}
\centering
\includegraphics[height=3.9in,width=6.5in]{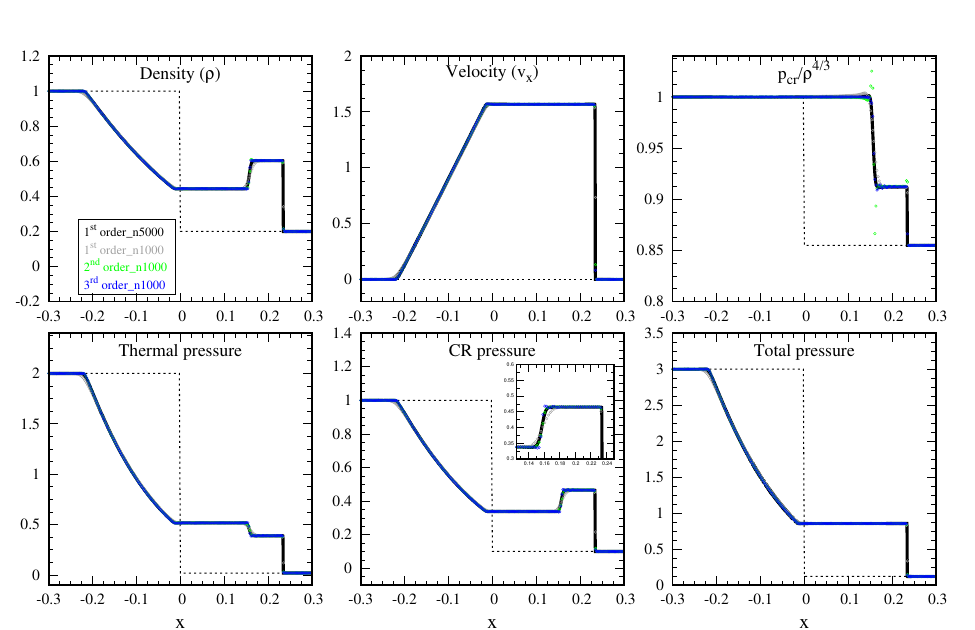}
\caption{Snapshot of various profiles from the two-fluid shock tube A (see Table \ref{tab:testproblems}) obtained using the \emph{Et+Ecr (Unsplit-pdv)} method. Black dashed curves show the profiles at $t=0$ and circles at $t=0.1$. Different colours represent results of different numerical schemes where `$\rm n$' denotes the number of grid points. All runs are performed using the CFL number $0.6$ (the results do not seem to depend on CFL numbers; see Appendix \ref{app:B}) and {\it without} fixing $p_{\rm cr}$ across the shock. The figure shows that the results do not depend on numerics, i.e., higher order solvers reproduce results identical to high resolution $1^{\rm st}$ order solvers.}
\label{fig:sodcr1}
\end{figure*}
\citet{Kudoh2016} showed that the amplitude of the spurious waves in Fig. \ref{fig:pbalance} (for the \emph{Et+Scr} method) can be reduced by increasing the spatial resolution or including additional numerical diffusion. However, both options are computationally expensive. 
 
\subsection{Shock tube A}\label{subsec:stA}
%
%
The initial condition for this test consists of two constant states separated by a discontinuity placed at $x=0$.
Left and right states are defined as $(\rho, v_{\rm x,} p_{\rm g}, p_{\rm cr})_{\rm L}=(1,0,2,1)$ and $(\rho, v_{\rm x,}, p_{\rm g}, p_{\rm cr})_{\rm R}=(0.2,0, 0.02,0.1)$ respectively. In all cases we employ $1000$ equally spaced zones. The set-up for this problem is identical to \citet{Kudoh2016}. 

Fig. \ref{fig:sodcr1_alldiff} shows fluid variable profiles at $t=0$ (black dashed curves) and $t=0.1$ (circles)  where different colours correspond to the four methods discussed in section \ref{sec:method} (also see section \ref{subsubsec:Pmod}). The solutions, obtained using $1^{\rm st}$-order reconstruction, are not identical.
Blue and green curves - representing the \emph{Eg+Ecr (OpSplit-pdv)} and \emph{Et+Ecr (Unsplit-pdv)} methods respectively - look similar although, as we shall see in Fig. \ref{fig:sodcr1zoom}, the solution obtained with the \emph{OpSplit} method depends on the details of spatial reconstruction and time-stepping.

We now turn our attention to the top rightmost panel of Fig. \ref{fig:sodcr1_alldiff} showing the CR entropy profile ($p_{\rm cr}/\rho^{4/3}$). In the two-fluid CR-HD model, it is sometimes assumed that CRs are adiabatically compressed in the post-shock region (\citealt{Pfrommer2006}). This assumption also helps to obtain an analytical solution of the shock tube problem. In numerical simulations this assumption is not automatically fulfilled. In fact, we find that CRs are adiabatically compressed only for the \emph{Et+Scr} method which is constructed to do so in the first place. In general, the post-shock solution depends on the method one uses.

As noted before, the computed results (for all methods except  \emph{unsplit-pdv} and  \emph{Et+Scr}) depend on numerical details such as reconstruction and the CFL number. While the \emph{Et+Scr} method is expected to yield robust results at shocks, a constant CR entropy across shocks is not physically justified as discussed earlier. Moreover, this method is unable to maintain pressure balance across a contact discontinuity as shown in section \ref{subsubsec:Pmod}.

Fig. \ref{fig:sodcr1zoom} shows the zoomed-in CR entropy profile from the shock tube problem using four different methods. In each panel, black, green and blue curves correspond to solutions obtained using $1^{\rm st}$, $2^{\rm nd}$ and $3^{\rm rd}$-order schemes. The top left panels indicate that the results are not unique: the profile obtained using the \emph{Et+Ecr} (\emph{Unsplit-vdp}, first panel) and \emph{Eg+Ecr} (\emph{OpSplit-pdv}, second panel) methods depend on the choice of spatial reconstruction. On the contrary, the \emph{Et+Scr} and \emph{Et+Ecr  (Unsplit-pdv)} schemes (shown in the $3^{\rm rd}$ and $4^{\rm th}$ columns) are in better agreement\footnote{We have also tested the HLLI solver of \citet{Dumbser2016} on this problem and find non-uniqueness of the solution.}.

We now consider the bottom panels of Fig. \ref{fig:sodcr1zoom} which, on the other hand, exhibit very similar profiles and have been obtained by fixing the CR pressure behind the shock, as explained in the following. First, a shock detection algorithm (section \ref{subsec:shockdetect}) is employed to identify shocked zones. Then we redistribute the shocked zone thermal and CR energies using the parameter
\begin{equation} \label{eq:shk}
\epsilon_{\rm shock} = \frac{e_{\rm cr}}{e_{\rm cr}+e_{\rm th}}
     = \frac{w_{\rm cr}(\gamma_{\rm g}-1)}
       {[\gamma_{\rm cr}-1 + w_{\rm cr}(\gamma_{\rm g}-\gamma_{\rm cr})]}  \,,
\end{equation} 
where
\begin{equation}
 \label{eq:shk2}
  w_{\rm cr} = \frac{p_{\rm cr}}{(p_{\rm g}+p_{\rm cr})} \,.
\end{equation} 
The redistribution of energy among CRs and the thermal fluid does not change the total energy (in particular $e_{\rm cr}+e_{\rm th}$) of the cell. In the bottom panels of Fig. \ref{fig:sodcr1zoom}, we have used $w_{\rm cr}=0.5$ to obtain the solutions. While using a parameter $w_{\rm cr}$ to fix the post-shock CR pressure, we notice that the post-shock conditions mainly depend on the total pressure of the shocked-zone nearest to the downstream. Thus, we are able to obtain identical numerical solutions irrespective of the method by imposing a sub-grid closure for CRs at shocks (see e.g., \citealt{Caprioli2014a}, \citealt{Mignone2018}). In section \ref{subsec:closure}, we have discussed some other possible equation-of-states (EoS) which can be used. Although a sub-grid closure is recommended, one would still prefer an implementation that is insensitive to the numerical details. Therefore, we recommend the `\emph{Et+Ecr (Unsplit-pdv)}' method.

Fig. \ref{fig:sodcr1} shows the robustness of the `\emph{Et+Ecr (Unsplit-pdv)}' method for different spatial reconstructions and time stepping (Appendix \ref{app:B} shows insensitivity of this method on the CFL number). As a reference solution, we have performed a simulation using an extremely high spatial resolution. The snapshot of various profiles at $t=0.1$ are shown. Four different solid curves, i.e., black/grey ($1^{\rm st}$ order), green ($2^{\rm nd}$ order), and blue ($3^{\rm rd}$ order) curves show the solutions with different spatial reconstruction. Since the profiles are identical, our implementation of the `\emph{Et+Ecr (Unsplit-pdv)}' method is numerically robust.

Summarizing our results, we conclude therefore that: (i) the solution of the two-fluid equations in the post-shock region depends on the choice of numerical methods, (ii) all methods give identical results only when the CR pressure behind the shock is imposed `by hand' (e.g., using a  subgrid closure at shocks \BLACK as given in Eq. \ref{eq:shk}), (iii) the results from our method \emph{Et+Ecr (Unsplit-pdv)} (Eqns. \ref{eq:v} and \ref{eq:pcr}) numerically robust even in absence of subgrid closure at shocks. For these reasons we discuss, from the next section onwards, the results from our best performing method, \quotes{\emph{Et + Ecr (Unsplit-pdv)}}.

\begin{figure*}
\centering
\includegraphics[height=2.25in,width=6.5in]{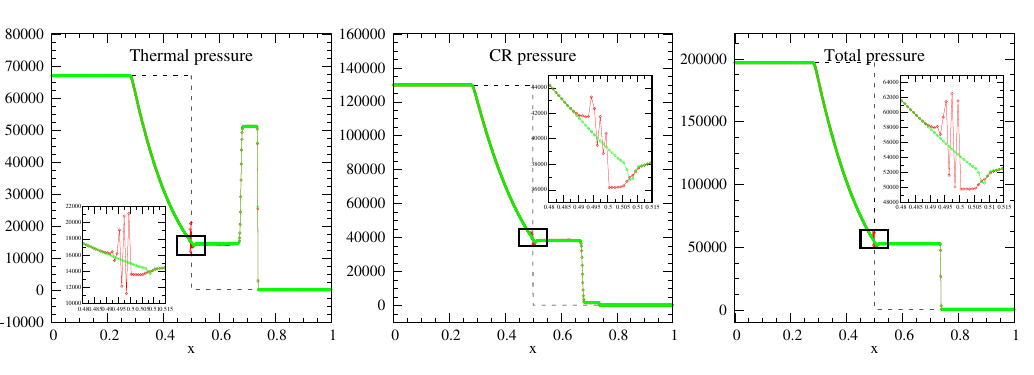}
\vspace{-2em}
\caption{Snapshot of various profiles at $t=10^{-4}$ from the shock tube B (problem $3$ in Table \ref{tab:testproblems}). Solutions are obtained by using \emph{Et+Ecr (Unsplit-pdv)} method. Two different colours, green and red, show profiles with and without altering the signal speed (Eqs. \ref{eq:s_um} and \ref{eq:s_m}). The sub-plots display the zoomed-in view of the black square. The figure shows that for green curves in which $\phi =1.1$ (Eq. \ref{eq:s_m}), the results are smooth.}
\label{fig:sodcr2}
\end{figure*}
\BLACK
\subsection{Shock tube B} \label{subsubsec:stB}
%
We now want to assess the sensitivity of the left- and right- going wave speed estimates in our method. Such estimates are typically required to calculate both the Riemann flux and the cell interface state of the conservative variables for HLL-type solvers. In order to check this, we set up the problem $3$ in Table \ref{tab:testproblems}, which leads to the formation of a right-going shock and left-going rarefaction waves.

In the literature, various choices are available for the left- and right- going wave speeds. (see e.g., Chapter $10$ in \citealt{Toro2009}). 
The simplest choice is the direct wave speed estimates suggested by \citet{Davis1988}, yielding (using the same notations of Fig. \ref{fig:nom}):
\begin{equation} \label{eq:s_um}
\begin{array}{l}
S_{{\rm R}, i+1/2} = \max (v^{+}_{i} + a^{+}_{i}, v^{-}_{i+1} + a^{-}_{i+1})\,, 
    \\ \noalign{\medskip} 
S_{{\rm L}, i+1/2} = \min (v^{+}_{i} - a^{+}_{i}, v^{-}_{i+1} - a^{-}_{i+1})\, .
\end{array}
\end{equation}
The present implementation of the \emph{Et+Ecr (Unsplit-pdv)} shows a spurious feature near the opening of the rarefaction wave (red curves in Fig. \ref{fig:sodcr2}).
We address this issue by adjusting the spatial width of the Riemann fan through a redefinition of the right- and left- wave speeds as
\begin{equation} \label{eq:s_m}
\begin{array}{l}
\tilde{S}_{R, i+\HALF} = \max (v^{+}_{i}   + \phi\,a^{+}_{i}, 
                               v^{-}_{i+1} + \phi\,a^{-}_{i+1})
                                 \\ \noalign{\medskip}
\tilde{S}_{L, i+\HALF} = \min (v^{+}_{i}   - \phi\, a^{+}_{i}, 
                               v^{-}_{i+1} - \phi\, a^{-}_{i+1})\, ,
\end{array}
\end{equation}
respectively, and choose ${\phi}  = 1.1$. Note that, when ${\phi} =1$ then Eq. (\ref{eq:s_m}) reduces to Eq. (\ref{eq:s_um}). 
This trick basically increases the width of the mixed state, which becomes $2\,\phi\, a$ at $t+\Delta t$, instead of $2\, a$ (here $a=(a_{\rm g}^2 + a_{\rm cr}^2)^{\HALF}$ is the sound speed of composite fluid). Increasing the width of the mixed state is justified by the fact that during the solution of the Riemann problem we do not include the coupling term which changes the effective signal speed (see e.g., Appendix \ref{app:A}). 
Although in the \emph{Unsplit-pdv} method we update the CR energy equation (Eq. \ref{eq:ecrup}) in a single step, the required sound wave speed may lie between $[a_{\rm g}^2 + a_{\rm cr}^2]^{\HALF}$ and $[a_{\rm g}^2 + (\gamma_{\rm g}/\gamma_{\rm cr})a_{\rm cr}^2]^{\HALF}$. Note that, the factor $\phi$ introduced here is obtained from our numerical experiments, with no rigorous proof. 

Fig. \ref{fig:sodcr2} shows different pressure profiles at $t=10^{-4}$ obtained with $\phi=1$ (red) and $\phi=1.1$ (green). For both solutions, in the post-shock region, the ratio $p_{\rm cr}/(p_{\rm cr}+p_{\rm g})\simeq 0.03$. Solutions are smooth only for  $\phi=1.1$, implying the necessity of a slightly higher wave speed estimate in order to obtain a robust solution.

\subsection{Sedov-Taylor Blast Wave} \label{subsec:multiD}
%
%
\begin{figure*}
\centering
\includegraphics[height=3.5in,width=6.9in]{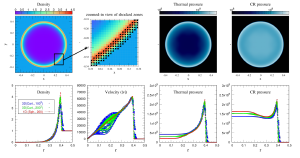}
\vspace{-2em}
\caption{Blastwave with CRs. Top panels show coloured maps of density, thermal and CR pressures from our $3$D run ($N_{\rm x}N_{\rm y}N_{\rm z} = 200^{3}$ grids) at $t=2\times 10^{-6}$ (in code units) in the $z=0$ plane. In the bottom panels, we compare radial profiles from the $1$D spherical (red) case with scatter plots of the $3$D Cartesian run using $100^3$ (blue) and $200^3$ (green) grid zones -- showing that $3$D profiles approach $1$D for a fine grid spacing. In the velocity plot ($2^{\rm nd}$ panel, bottom) we notice that the fluid velocity in the $3$D runs show some deviations from the $1$D spherical run. We find that these deviations can be reduced by setting a large $r_{\rm inj}$ so that the energy injection region at $t=0$ is close to spherical. Shocked zones (where CRs are injected using $w_{\rm cr}=0.5$) are shown by black solid squares in the $2^{\rm nd}$ panel, top row.}
\label{fig:blastcr}
\end{figure*}

We have extended our implementation of CR-HD equations to multi-dimensions and different geometries in the \PLUTO code. We perform the standard blast wave problem in $3$D Cartesian and $1$D spherical geometries. The numerical set-up is discussed below.

At $t=0$, we create high pressure in a small region by setting $p_{\rm g} = (\gamma_{\rm g}-1) E/\Delta V$ where $\Delta V = 4/3\pi\, r_{\rm inj}^{3}$ is the small volume, $r_{\rm inj} = 0.01$ (in code units), and $E=10^{51}$ erg. 
In the rest of the computational domain, density, velocity, and pressures are set to $1$, $0$, and $60$ respectively. We normalize density, velocity, and length to $\rho_u = m{\rm_{H}\,cm^{-3}}$ ($1.67 \times 10^{-24}$ g cm$^{-3}$),  $v_u = 10^{5}\,{\rm cm\, s^{-1}}$ and $L_u = 3.086\times 10^{18}\,{\rm cm}$ ($1$ parsec) respectively. 
For $1$D spherical geometry, the inner radial boundary at $r=0.001$ is set to be  reflective while for $3$D Cartesian, we set all boundaries to outflow. 
For the 3D run, we employ $200$ equally spaced zones along $x, y$, and $z$ directions, i.e.,  $N_{\rm x}N_{\rm y}N_{\rm z}=200^3$ (but we also perform a low resolution run using $N_{\rm x}N_{\rm y}N_{\rm z}=100^3$ to check convergence), and for the $1$D run we use $200$ equally spaced zones. 
Simulations are performed using a CFL number $C_{\rm a} = 0.2$ for both geometries.
Since we do not inject mass initially, the Sedov-Taylor phase starts right at the beginning of the shock evolution. The initial injection is purely thermal, but we inject CRs at the shock as follows.

First, we identify shocked zones using the detection algorithm (see section \ref{subsec:shockdetect}) and then inject CRs in these zones using the parameter $w_{\rm cr}$ (Eq. \ref{eq:shk}) as done previously in 1D. 
We set $w_{\rm cr}=0.5$, i.e., equipartition between thermal and CR pressures in the post-shock zones. The results from $3$D (Cartesian) and $1$D (spherical) runs are shown in Fig. \ref{fig:blastcr}. In the top panels we show the snapshots of density, thermal and CR pressures in the $z=0$ plane  at $t = 2\times 10^{-6}$ (code unit) obtained from our $3$D run. The second panel (top row) shows the zoomed-in view of the shocked zones where CR energy is injected. In the bottom panels we compare the results between the $3$D case and the $1$D runs, showing good agreement. We, therefore, conclude that our implementation (i.e., \emph{Et+Ecr (Unsplit-pdv)} method) is well suited for multi-dimensional calculations.

\section{Discussion}\label{sec:discussions} 
%
%
The results from the previous section have shown that the numerical solution of the two-fluid equations depends on the discretization of the coupling term and that the results become insensitive to the choice of the method only when the post-shock CR pressure is fixed by the parameter $w_{\rm cr}$ (Fig. \ref{fig:sodcr1zoom}), which provides a possible closure to the problem discussed in section \ref{subsec:closure}. Below we further discuss to what extent the CR diffusion term that we have neglected until now can affect our conclusions, and also present the limitations and broader implications of this work.  

\subsection{Shock structure with CR diffusion}
Several works on two-fluid shocks (\citealt{Drury1981,Becker2001}) have considered the impact of CR diffusion on the shock structure. 
A non-negligible CR diffusion implies that the CR pressure (unlike the gas pressure) remains continuous across the shock transition (length scale $\sim \kappa_{\rm cr}/v_{\rm 1}$). Since the spatial width of the shock is now broadened, some additional equations can be formulated between  far- upstream and downstream of a shock, which are known as precursor-EoSs (\citealt{Drury1981,Voelk1984,Jun1994}). 
The precursor-EoS connects the upstream and downstream and thereby allows to investigate the shock structure, and it can be seen as a closure to two-fluid equations. Although this method has been promising to study shock modification due to CR acceleration, some features are worth highlighting.

In CR-HD, two distinct shock structures are possible: a discontinuous, gas-mediated sub-shock with CRs diffusing ahead of the shock; and a  CR-dominated smooth \quotes{shock} (across which the gas properties vary smoothly) for large Mach numbers, i.e., $\gtrsim 12$, see Fig. 2 in \citealt{Becker2001}, or the right panels in Fig. 5 of \citealt{Gupta2018a} for an astrophysical realization in a wind-driven shock. In the latter case, most of the upstream kinetic energy flux goes into CR acceleration rather than heating the thermal plasma, and the astrophysical implications of this are enormous (see e.g., \citealt{Gupta2018b}). However, the physical existence of such shocks is yet to be established by kinetic plasma simulations. In some parameter space, multiple solutions are predicted for the same set of upstream conditions and the results are found to be sensitive to the upstream variables close to the shock surface  (for a brief discussion, see section $6$ in \citealt{Becker2001}). 

Apart from the multiplicity (non-unique) of the post-shock solutions, the choice of the CR diffusion coefficient is crucial. Near a shock, magnetic fluctuations due the plasma instabilities (\citealt{Kulsrud1969, Bell2004}) can significanly modify $\kappa_{\rm cr}$ from those used in large scale flows (\citealt{Schroer2020}). Therefore although in theory, CR diffusion allows to find shock structure, the results on the shock structure are inconclusive and further investigation is needed. The results from kinetic studies, irrespective of whether a globally smooth shock forms or not, can always be implemented as a subgrid EoS.

\subsection{Limitations and broader implications}
Non-unique numerical results at the shocks that have been discussed in this work is ultimately connected to the missing microphysics in fluid approximation (\citealt{Hoyle1960}). 
Both fluid and kinetic studies showed that CRs drive a zoo of plasma instabilities that are responsible for CR acceleration and eventualy build-up the post-shock CR pressure  (\citealt{Dorfi1985,Drury1986,Ryu1993}; for a review see \citealt{Marcowith2016}). Since the majority of these instabilities occurring on the CR Larmor radius scale depend on the CR energy flux, neglecting them 
far away from the shock is justified as long as their effects are incorporated in the subgrid EoS connecting the far-upstream and downstream regions. However, to obtain a detailed shock structure in fluid theory, we must include a self-consistent (derived from kinetic considerations) CR diffusion. 

Note that the CR injection model presented in this work (see e.g., Fig. \ref{fig:blastcr}) is different from those suggesting injection at the gas sub-shocks (see e.g., \citealt{Falle1987,Kang1990,Saito2013}). The sub-shock injection still relies on the fluid approximation and the results can be affected by the choice of the diffusion coefficient and other missing kinetic processes at shocks. Moreover, injection at the gas sub-shock requires resolving the shock transition region (\citealt{Tsung2020}) which is computationally infeasible in galaxy-scale simulations. From this perspective, our method may seem a simplification, but we believe that in absence of a self-consistent model that takes into account the necessary shock microphysics, the proposed method can give better control on shock solutions and their impact on the large scale flow. 

Although we recommend using such closures at shocks, it is worth mentioning that this requires shock identification on the computational grid. 
We suggest the \quotes{\emph{Et+Ecr (Unsplit-pdv)}} method (section \ref{subsec:method2}) for a robust numerical solution, irrespective of the use of a physically motivated subgrid closure at shocks.

The main limitation of our method is that it does not allow investigating the time evolution of shock structure self-consistently. This is not the aim of studies large scale impact of CRs (length scale $\gg$ shock transition width). By defining the CR injection as a function of time, the shock evolution may be inspected. A physically reliable value of $w_{\rm cr}$ (or an equivalent parameter; e.g., see section \ref{subsec:closure} depends on upstream parameters (e.g., Mach number and magnetic field orientation w.r.t. shock normal) and it needs to be prescribed using Particle-In-Cell simulations (see e.g., \citealt{Caprioli2014a}). In MHD, the injection of CRs will also depend on the nature of the shock (e.g., fast versus slow) and detailed calibration with kinetic plasma simulations will be needed (\citealt{Bret2020}).

\section{Summary} \label{sec:conclu}
In this work we have shown that the two-fluid CR-HD model suffers from a non-uniqueness problem at shocks. It contains three conservation laws and one additional equation (for the CR pressure) which cannot be cast in a satisfactory conservative form and causes difficulties in its numerical implementation. A unique shock jump condition is possible only if one makes an additional assumption. The steady-state shock structures can be predicted by assuming a suitable downstream CR pressure/energy (as discussed in section \ref{subsec:closure}). There is a degeneracy between the gas and CR pressures (see last paragraph of section \ref{sec:basics}) because of which the solutions may depend on numerics. Without fixing the fraction of upstream energy transformed into CR energy and simply relying on the numerical discretization of the non-conservative exchange term involving derivatives $p_{\rm cr} \nabla \cdot {\bf v}$ (or equivalently, ${\bf v} \cdot \nabla p_{\rm cr}$), makes the solutions of the two-fluid equations across shocks depend on the details of numerics (e.g., spatial reconstruction, time-stepping, and even the CFL number). In this work, we have investigated numerical implementation of the two-fluid CR-HD equations as applied to large scale simulations. Our findings are summarized as follows:

\begin{itemize}
\item Numerical solution of the two-fluid equations depends on implementation of the coupling term ($p_{\rm cr}{\nabla \cdot \bf{v}}$ or ${\bf v \cdot \nabla}p_{\rm cr}$). We show that the different discretizations do not show an identical solution (Figs. \ref{fig:sodcr1_alldiff} and  \ref{fig:sodcr1zoom}). This is because of the non-negligible and non-unique contribution of the source term involving 
a derivative at the shock.

\item We suggest a method (\quotes{\emph{Et+Ecr (Unsplit-pdv)}}, section \ref{subsec:method2}) for which the solutions are robust to the choice of spatial reconstruction, time-stepping, and the CFL number (see Fig. \ref{fig:sodcr1}). In order to ensure that the characteristic speed remains within the fastest signal propagation speed, this method demands a slightly higher signal speed than the standard estimate of the two-fluid sound speed (Fig. \ref{fig:sodcr2}). 

\item We show that all methods give an identical solution only when the CR pressure across the shock is fixed {\it by an imposed equation of state}. This can be done, for example, by specifying CR pressure in the shocked zones (as done in Fig. \ref{fig:sodcr1zoom}; also see section \ref{subsec:closure}). A physically realistic implementation of the CR pressure across a shock is possible by calibrating with kinetic simulations using different upstream parameters. In this approach, a shock must be properly identified. We suggest the \quotes{\emph{Et+Ecr (Unsplit-pdv)}} method for a robust solution irrespective of whether a subgrid closure is used at shocks.
\end{itemize}

In summary, this work highlights the critical aspects of the two-fluid CR-HD equations. Although here we have not discussed the CR-MHD system, these problems are also present there. The implementation of CR-MHD will be discussed in a future work. 
 
\section*{Acknowledgements}
We thank Dinshaw Balsara, Damiano Caprioli, Michael Dumbser, Yuki Kudoh, and Christoph Pfrommer for helpful discussions. SG acknowledges CSIR, India for the SPM Fellowship. SG thanks the postdoctoral research program of the University of Chicago, where some of this work was done to address referee's suggestions. PS thanks the Humboldt Foundation for supporting his sabbatical at MPA where some of this work was done. PS acknowledges the Department of Science and Technology for a Swarnajayanti Fellowship (DST/SJF/PSA-03/2016-17). We thank our reviewer, Luke Drury, for constructive comments on our manuscript.

\section*{Data Availability}
The data and code associated with the paper can be made available on request to the corresponding author.

\bibliographystyle{mnras}
\bibliography{reference-list}

\onecolumn 
\appendix
\vspace{-1em}
\section{Details of various EoSs} \label{app:eos_detail}
%
Here we present the detailed calculations for finding shock solution for a given EOS outlined in section \ref{subsec:closure}.
\vspace{-1em}
\subsection{w$_{\rm cr}$-EoS}\label{app:wcr}
Using Eq. (\ref{eq:wcr}), we substitute $\mathcal{P}_{\rm g,2}$ in Eqs. (\ref{eq:mom_c2}) and (\ref{eq:engy_c2}). After combining them, we obtain a quadratic equation:
\begin{eqnarray}\label{eq:solwcrR}
\mathcal{R}^2\,\mathcal{A}- \mathcal{R}\, \mathcal{B}+ \mathcal{C}=0\, , \ {\rm where}
\end{eqnarray} 
\begin{eqnarray}
\mathcal{A} &=& \left\{(\gamma_{\rm g} +\gamma_{\rm cr}\mathcal{P}_{\rm cr,1})\frac{\mathcal{M}^2_{\rm 1}}{2} +\frac{\gamma_{\rm g}}{\gamma_{\rm g}-1} + \frac{\gamma_{\rm cr}}{\gamma_{\rm cr}-1} \mathcal{P}_{\rm cr,1}   \right\}\\
\mathcal{B} &=& \left\{\frac{\gamma_{\rm g}}{\gamma_{\rm g}-1}(1-w_{\rm cr})+\frac{\gamma_{\rm cr}}{\gamma_{\rm cr}-1}w_{\rm cr}\right\}\left\{(\gamma_{\rm g} +\gamma_{\rm cr}\mathcal{P}_{\rm cr,1})\mathcal{M}^2_{\rm 1}+(1+\mathcal{P}_{\rm cr, 1})\right\}\\
\mathcal{C} & = & \left\{\frac{\gamma_{\rm g}}{\gamma_{\rm g}-1}(1-w_{\rm cr})+\frac{\gamma_{\rm cr}}{\gamma_{\rm cr}-1}w_{\rm cr} -\frac{1}{2}\right\}\left\{(\gamma_{\rm g} +\gamma_{\rm cr}\mathcal{P}_{\rm cr,1})\mathcal{M}^2_{\rm 1}\right\} \  .
\end{eqnarray}
Note that $w_{\rm cr}$ is a parameter specified in the downstream, and $\mathcal{P}_{\rm cr, 1}$ (Eq. \ref{eq:pcr1_nom}) and $\mathcal{M}_{\rm 1}(\mathcal{M}_{\rm cr, 1},\mathcal{M}_{\rm cr, 1})$ are upstream parameters. The solutions of Eq. (\ref{eq:solwcrR}) is given by
\begin{eqnarray}\label{eq:sol_wcrR}
\mathcal{R}_{\rm \pm}   = \frac{\mathcal{B}\pm (\mathcal{B}^2-4\mathcal{A}\mathcal{C})^{\HALF}}{2 \mathcal{A}} \ ,
\end{eqnarray}
Since $\mathcal{R}_{\rm -}\leq 1$, $\mathcal{R}_{\rm +}$ represents the physical solution and its values as a function of upsteam Mach numbers for four different $w_{\rm cr}$ are shown in  Fig \ref{fig:eos_wcr}. The figure indicates that the compression ratio increases with $w_{\rm cr}$. It can also be shown analytically:
\begin{eqnarray} \label{Eq:wcr_R_Mhigh}
 \lim_{\mathcal{M}_{\rm 1} \to\infty} \mathcal{R}_{\rm +}  = \frac{\gamma_{\rm g}+1}{\gamma_{\rm g}-1} + 2\,w_{\rm cr}\left(\frac{\gamma_{\rm cr}}{\gamma_{\rm cr}-1}-\frac{\gamma_{\rm g}}{\gamma_{\rm g}-1}\right) \, .
\end{eqnarray}
Taking $\gamma_{\rm g}=5/3$ and $\gamma_{\rm cr}=4/3$, we obtain $\mathcal{R}_{\rm +} = 4+3  w_{\rm cr}$, which shows that the compression ratio approaches $7$ when $w_{\rm cr}\rightarrow 1$. Note that, CR diffusion can increase the compression ratio even for a smaller $w_{\rm cr}$ due to diffusive loss of CR energy (section $4.2.1$ in \citealt{Gupta2018a}).

\vspace{-1em}
\subsection{$\epsilon_{\rm cr}$-EoS}\label{app:epscr}
\begin{figure}
\centering
\includegraphics[height=1.9in,width=5.2in]{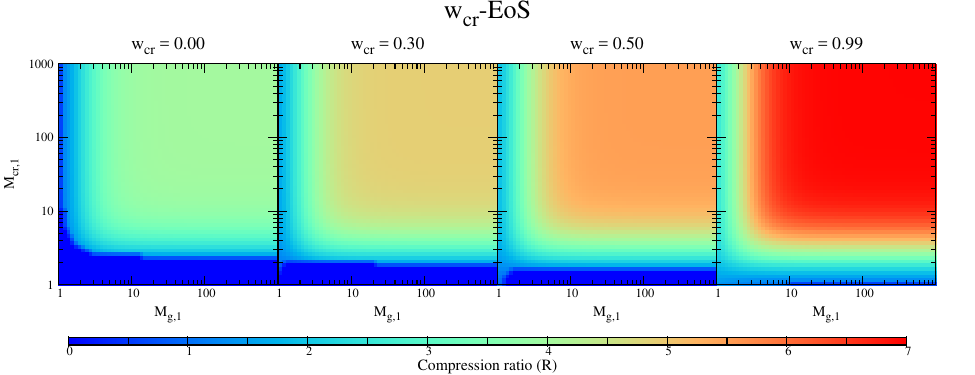}
\caption{Compression ratio ($\mathcal{R}$) as a function of upstream gas ($\mathcal{M}_{\rm g, 1}$) and CR ($\mathcal{M}_{\rm cr, 1}$) Mach numbers for four different $w_{\rm cr}$. The left-most panel represents a special case, $w_{\rm cr} = 0$ (corresponding to a downstream CR pressure $p_{\rm cr, 2}\rightarrow 0$), which can be considered as a one-fluid HD model in the limit $\mathcal{M}_{\rm cr, 1}\rightarrow \infty$ (corresponding to $p_{\rm cr, 1}\rightarrow 0$). For a large $\mathcal{M}_{\rm g, 1}$, this panel shows $\mathcal{R}\rightarrow 4$ (see the colour palette), as expected from the Rankine-Hugoniot shock-jump condition. As expected, a larger $w_{\rm cr}$ gives a higher compression ratio.}
\label{fig:eos_wcr}
\centering
\includegraphics[height=1.9in,width=5.2in]{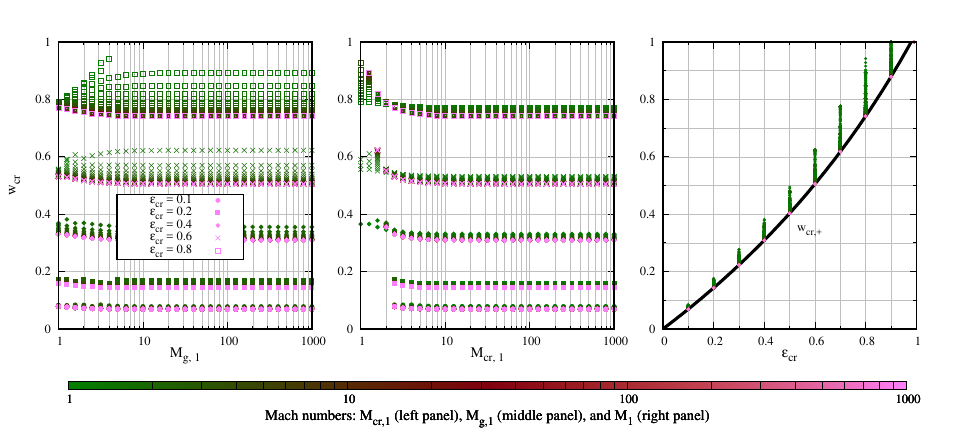}
\caption{Dependence of $w_{\rm cr}$ (Eq. \ref{eq:wcr}) on the upstream Mach numbers and $\epsilon_{\rm cr}$ (Eq. \ref{eq:pcr1}). Each symbol in left and middle panels represents solution of  $w_{\rm cr}$ for a given $\epsilon_{\rm cr}$, where the colours represent the upstream CR ($\mathcal{M}_{\rm cr, 1}$) and gas ($\mathcal{M}_{\rm g, 1}$) Mach numbers respectively. These two panels show that for a high Mach number flow the parameter $w_{\rm cr}$ mainly depends on $\epsilon_{\rm cr}$. The right panel shows the dependence of $w_{\rm cr}$ on $\epsilon_{\rm cr}$. The black solid curve shows the analytic solution (Eq. \ref{eq:wcr_anaMinf2}).}
\label{fig:rela_wcr_ecr}
\end{figure}

In this case, we follow a similar approach as done in Appendix \ref{app:wcr}. We first normalize the gas and CR pressures of Eq. (\ref{eq:pcr1}) w.r.t. the upstream gas pressure and find
\begin{eqnarray}\label{eq:pcr2} 
\mathcal{P}_{\rm cr,2}&= & \epsilon_{\rm cr}\,\frac{\gamma_{\rm cr}-1}{\gamma_{\rm cr}} \mathcal{R}\left(\gamma_{\rm g} +\gamma_{\rm cr}\mathcal{P}_{\rm cr,1}\right)\left\{\frac{\mathcal{M}_{\rm 1}^2}{2} + \frac{1}{\gamma_{\rm g} +\gamma_{\rm cr}\mathcal{P}_{\rm cr,1}}\left( \frac{\gamma_{\rm g}}{\gamma_{\rm g}-1} + \frac{\gamma_{\rm cr}}{\gamma_{\rm cr}-1} \mathcal{P}_{\rm cr,1}\right)\right\} \ .
\end{eqnarray}
Then we substitute $\mathcal{P}_{\rm cr, 2}$ (Eq. \ref{eq:pcr2}) in Eqs. (\ref{eq:mom_c2}) and (\ref{eq:engy_c2}), which results in a quadratic equation identical to Eq (\ref{eq:solwcrR}), where 
\begin{eqnarray}
\mathcal{A} &=&(\gamma_{\rm g} +\gamma_{\rm cr}\mathcal{P}_{\rm cr,1})
\mathcal{M}^2_{\rm 1}\left\{\frac{\gamma_{\rm g}-1}{2\gamma_{\rm g}} - \frac{\gamma_{\rm cr}-1}{2\gamma_{\rm cr}}\epsilon_{\rm cr}\left(\frac{\gamma_{\rm g}-1}{\gamma_{\rm cr}-1}\frac{\gamma_{\rm cr}}{\gamma_{\rm g}}-1\right)\right\} + 1 +
\epsilon_{\rm cr}\left(\frac{\gamma_{\rm cr}-1}{\gamma_{\rm g}-1}\frac{\gamma_{\rm g}}{\gamma_{\rm cr}} - 1\right)  + \mathcal{P}_{\rm cr,1} \left\{(1-\epsilon_{\rm cr})\left(\frac{\gamma_{\rm g}-1}{\gamma_{\rm cr}-1}\frac{\gamma_{\rm cr}}{\gamma_{\rm g}}\right)+\epsilon_{\rm cr}\right\} 
\\
\mathcal{B} &=& \left\{(\gamma_{\rm g} +\gamma_{\rm cr}\mathcal{P}_{\rm cr,1})\mathcal{M}^2_{\rm 1}+(1+\mathcal{P}_{\rm cr, 1})\right\}\\
\mathcal{C} &=&  (\gamma_{\rm g} +\gamma_{\rm cr}\mathcal{P}_{\rm cr,1})\mathcal{M}^2_{\rm 1} \left(\frac{\gamma_{\rm g}+1}{2\gamma_{\rm g}}\right)
\end{eqnarray}
Therefore, the solutions can be found by solving Eq. (\ref{eq:sol_wcrR}). We find that the qualitative nature of the solutions remains similar to Fig. \ref{fig:eos_wcr}, i.e., the compression ratio increases with $\epsilon_{\rm cr}$.
\subsubsection{Relation between w$_{\rm cr}$ and $\epsilon_{\rm cr}$}\label{app:rela_wcrecr}
The relation between w$_{\rm cr}$-EoS and $\epsilon_{\rm cr}$-EoS can be obtained as follows. From Eq. (\ref{eq:wcr}), we note that the value of $w_{\rm cr}$ depends on the downstream CR and gas pressures. In $\epsilon_{\rm cr}$-EOS, the downstream CR pressure ($\mathcal{P}_{\rm cr,2}$) is already parameterized by $\epsilon_{\rm cr}$, which is given in Eq. (\ref{eq:pcr2}). The remaining quantity, $\mathcal{P}_{\rm g,2}$, can be obtained using the momentum conservation equation, Eq. (\ref{eq:mom_c2}), which gives
\begin{eqnarray}\label{eq:pg2}
\mathcal{P}_{\rm g, 2} = \left\{\left(\gamma_{\rm g} +\gamma_{\rm cr}\mathcal{P}_{\rm cr,1}\right)\mathcal{M}_{\rm 1}^2\left(1-\frac{1}{\mathcal{R}}\right)+1+\mathcal{P}_{\rm cr,1}\right\} - \mathcal{P}_{\rm cr,2}.
\end{eqnarray}
It is worth mentioning that the only constrain on the above equation is $\mathcal{P}_{\rm g, 2}> 0$, implying that the downstream CR pressure $\mathcal{P}_{\rm cr,2}< \left\{\left(\gamma_{\rm g} +\gamma_{\rm cr}\mathcal{P}_{\rm cr,1}\right)\mathcal{M}_{\rm 1}^2\left(1-\frac{1}{\mathcal{R}}\right)+1+\mathcal{P}_{\rm cr,1}\right\}$. Since we now have $\mathcal{P}_{\rm g,2}$ and $\mathcal{P}_{\rm cr,2}$, we can obtain $w_{\rm cr}$ as a function of $\mathcal{R}$, $\mathcal{M}_{\rm 1}$, $\mathcal{P}_{\rm cr, 1}$, and $\epsilon_{\rm cr}$ where $\mathcal{R}(\mathcal{M}_{\rm 1} , \mathcal{P}_{\rm cr, 1}, \epsilon_{\rm cr})$ can be calculated from the solution of Eq. (\ref{eq:sol_wcrR}). Therefore, the parameter $w_{\rm cr}$ becomes a function of $\mathcal{M}_{\rm g, 1}$, $\mathcal{M}_{\rm cr, 1}$, and $\epsilon_{\rm cr}$. The value of $w_{\rm cr}$ for various upstream Mach numbers and $\epsilon_{\rm cr}$ are shown in Fig. \ref{fig:rela_wcr_ecr}. In the limit $\mathcal{M}_{\rm 1}\rightarrow \infty$, it can be shown that
\begin{eqnarray}\label{eq:wcr_anaMinf2}
w_{\rm cr} = \frac{\epsilon_{\rm cr}}{8}\frac{\mathcal{R}_{\rm +}(w_{\rm cr})}{1-1/{\mathcal{R}_{\rm +}(w_{\rm cr})}} \rightarrow w_{\rm cr, \pm} =  \frac{4\left\{-3(1-\epsilon_{\rm cr})\pm (9+6\epsilon_{\rm cr})^{1/2}\right\}}{3(8-3\epsilon_{\rm cr})},
\end{eqnarray}
where we have taken $\mathcal{R}_{\rm +}$ from Eq. (\ref{Eq:wcr_R_Mhigh}) and the constrain $0\leq w_{\rm cr}\leq 1$ implies that $w_{\rm cr, +}$ is the physical solution.
\subsection{Adiabatic-EoS}\label{app:adia}
\begin{figure}
\centering
\includegraphics[height=1.9in,width=4in]{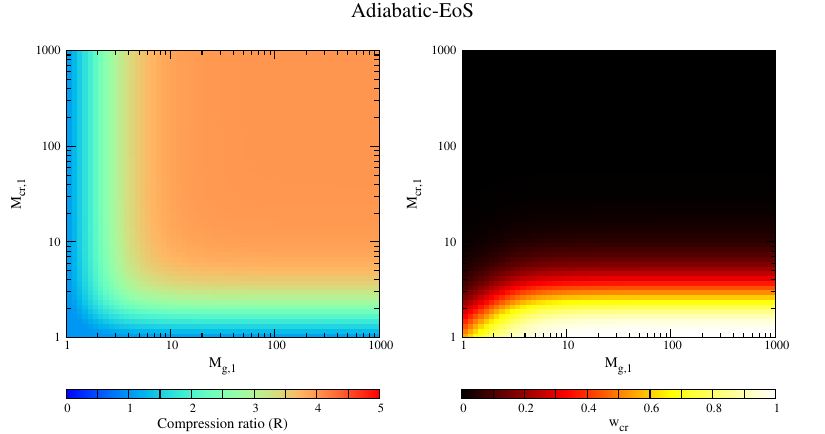}
\caption{Solution for an adiabatic EoS. [Left] Compression ratio ($\mathcal{R}$) as a function of upstream gas ($\mathcal{M}_{\rm g, 1}$) and CR ($\mathcal{M}_{\rm cr, 1}$) Mach numbers, which shows that when both $\mathcal{M}_{\rm g,1}$ and $\mathcal{M}_{\rm cr,1} >> 1$, $\mathcal{R}\rightarrow 4$. [Right] Dependences of $w_{\rm cr}$ on the upstream Mach numbers. It shows that for a large $\mathcal{M}_{\rm cr,1}$ (corresponding to a low upstream CR pressure), $w_{\rm cr}\rightarrow 0$. In other words, it indicates that if upstream CR pressure is very low compared to upstream gas pressure then the post-shock CR pressure is almost negligible compared to the gas pressure. For a large upstream CR pressure, the value of $w_{\rm cr}\rightarrow 1$, i.e., the downstream is mostly dominated by CR pressure.}
\label{fig:eos_adia}
\end{figure}

To find the compression ratio, we substitute $\mathcal{P}_{\rm cr, 2}$ from Eq. (\ref{eq:eos-adia}) to Eq. (\ref{eq:mom_c2}) and obtain $\mathcal{P}_{\rm g, 2}$ as a function of $\mathcal{R}$, $\mathcal{P}_{\rm cr,1}$, and $\mathcal{M}_{\rm 1}$. Replacing this $\mathcal{P}_{\rm g, 2}$ in Eq. (\ref{eq:engy_c2}) we obtain
\begin{eqnarray} \label{eq:sol-adia}
\mathcal{R}^{\gamma_{\rm cr}+1}\, \mathcal{A} - \mathcal{R}^2\, \mathcal{B} +  \mathcal{R}\, \mathcal{C} + \mathcal{D} = 0, \ {\rm where }
\end{eqnarray}
\begin{eqnarray}
\mathcal{A} &=& \left(\frac{\gamma_{\rm cr}}{\gamma_{\rm cr}-1}-\frac{\gamma_{\rm g}}{\gamma_{\rm g}-1} \right) \mathcal{P}_{\rm cr, 1}\nonumber \\
\mathcal{B} &=& \left\{(\gamma_{\rm g} +\gamma_{\rm cr}\mathcal{P}_{\rm cr,1})\frac{\mathcal{M}^2_{\rm 1}}{2} +\frac{\gamma_{\rm g}}{\gamma_{\rm g}-1} + \frac{\gamma_{\rm cr}}{\gamma_{\rm cr}-1} \mathcal{P}_{\rm cr,1}   \right\};\nonumber\\
\mathcal{C} &=& \frac{\gamma_{\rm g}}{\gamma_{\rm g}-1}\left\{\mathcal{M}_{\rm 1}^{2}(\gamma_{\rm g} +\gamma_{\rm cr}\mathcal{P}_{\rm cr,1})+1+\mathcal{P}_{\rm cr,1}\right\} \nonumber\\
\mathcal{D} &=& \mathcal{M}^2_{\rm 1} (\gamma_{\rm g} +\gamma_{\rm cr}\mathcal{P}_{\rm cr,1})\,\left(\frac{1}{2}-\frac{\gamma_{\rm g}}{\gamma_{\rm g}-1}\right).
\end{eqnarray}
Eq. (\ref{eq:sol-adia}) can be solved using a standard root-finding scheme (e.g., the Newton-Raphson method). After obtaining the solution for $\mathcal{R}$, we can calculate $w_{\rm cr}$ (Eq. \ref{eq:wcr}) and the results are shown in Fig. \ref{fig:eos_adia}. The left panel of Fig. \ref{fig:eos_adia} displays the compression ratio as a function of $\mathcal{M}_{\rm g, 1}$ and $\mathcal{M}_{\rm cr, 1}$, which shows that the compression ratio $\leq 4$. The right panel shows $w_{\rm cr}$ as a function of $\mathcal{M}_{\rm g, 1}$ and $\mathcal{M}_{\rm cr, 1}$. It shows that if  the upstream CR pressure is very small compared to the upstream ram pressure (i.e., when $\mathcal{M}_{\rm cr, 1}\gg 1$) then the post-shock CR pressure is almost negligible compared to the gas pressure. For a large upstream CR pressure (${\cal M}_{\rm cr, 1} \sim 1$), the value of $w_{\rm cr}\rightarrow 1$, i.e., the downstream is mostly dominated by CR pressure. However, even in this regime, the compression ratio $\leq 4$, which can be seen by comparing the left and right panels. These results are easy to understand intuitively because the effective Mach number $\mathcal{M_{\rm 1}}\sim 1$; i.e., the shock is too weak to show a noticeable change in CR pressure.
\BLACK
\section{Eigenvalues without the coupling term} \label{app:A}
\vspace{-0.5em}
To obtain the HLL flux at left- and right-interface of the computational cell, we need to provide an estimate of the signal speed to the solver. Since in the operator splitting method the solver does not have any information of the $p_{\rm cr} \nabla \cdot {\bf v}$ (or ${\bf v} \cdot\nabla p_{\rm cr}$) coupling term, the signal speed can be different from the actual speed of the complete system. This is illustrated as follows. Consider that the source term in Eq. (\ref{eq:gen}), $p_{\rm cr}\nabla\cdot {\bf v}$, is absent. In 1D Cartesian geometry, the Jacobian matrix without the coupling and diffusion terms is found to be
\begin{equation}
J = \begin{bmatrix} 
  0 & 1 & 0 & 0 
  \\
  \\
\frac{1}{2}(\gamma_{\rm g}-3) v^2 &(3-\gamma_{\rm g})\,v & (\gamma_{\rm g}-1) & (\gamma_{\rm cr}-\gamma_{\rm g}) 
\\
\\
\frac{1}{2}(\gamma_{\rm g}-3) v^3-v\left(\frac{a_{\rm g}^2}{\gamma_{\rm g}-1} +\frac{a_{\rm cr}^2}{\gamma_{\rm cr}-1}\right) & \frac{1}{2}(3-2\gamma_{\rm g}) v^2+\left(\frac{a_{\rm g}^2}{\gamma_{\rm g}-1} +\frac{a_{\rm cr}^2}{\gamma_{\rm cr}-1}\right) & \gamma_{\rm g} v & (\gamma_{\rm cr}-\gamma_{\rm g}) v
\\
\\
-\frac{v a_{\rm cr}^2}{\gamma_{\rm cr}(\gamma_{\rm cr}-1)} & \frac{a_{\rm cr}^2}{\gamma_{\rm cr}(\gamma_{\rm cr}-1)} & 0 & v
\end{bmatrix}
\end{equation}
Eigenvalues of this matrix are $\lambda=v-a_{\rm eff}, v, v, v+a_{\rm eff}$ where $a_{\rm eff}=[a^2_{\rm g} + (\gamma_{\rm g}/\gamma_{\rm cr})a^2_{\rm cr}]^{\HALF}$. The methods involving the $pdv$ source term give the same set of eigenvalues. As $\gamma_{\rm g}>\gamma_{\rm cr}$, we note that the effective propagation speed $a_{\rm eff}$ is slightly larger than the actual sound speed of the composite fluid, which is $[a^2_{\rm g} + a^2_{\rm cr}]^{\HALF}$. Without using the effective sound speed as an estimate of the maximum signal propagation speed, one may find spurious oscillations in the solution. Such an example has been discussed in section \ref{subsubsec:stB}. The effective sound speed ($a_{\rm eff}$) with vdp splitting is also different from $[a_{\rm g}^2+a_{\rm cr}^2]^{1/2}$, and we advice caution even in this case for the signal speed estimate.
\section{Dependences on the CFL number} \label{app:B}
\vspace{-0.5em}
Other than for the \emph{Et+Scr} method, the coupling term has to be implemented in the two-fluid equations. In order to implement this term, we have to choose $p_{\rm cr}$ and $v$ at the cell center/interface, which can be chosen in various ways. We find that all of these choices do not produce a unique/consistent result. To explicitly show this, we present the result from the shock tube A (problem $2$ in Table \ref{tab:testproblems}) using three possible combinations of $p_{\rm cr}$ and $v$ in Eq. (\ref{eq:ecrup}). These choices are given below.
\begin{eqnarray}
 {\rm  Possibility\, 1:} & & 
\begin{cases}
    v^{n}_{i-\HALF} = \frac{1}{2}(v^{+, n}_{\rm i-1}+v^{-, n}_{\rm i-1});\ v^{n}_{i+\HALF} = \frac{1}{2}(v^{+, n}_{\rm i}+v^{-, n}_{\rm i});\, p^{n}_{\rm cr, i} =\frac{1}{2}(p^{+,n}_{\rm cr, i}+p^{-,n}_{\rm cr, i});\ \, \ v^{n}_{\rm i}\geq 0\nonumber \nonumber \\ 
   v^{n}_{i-\HALF} = \frac{1}{2}(v^{+, n}_{\rm i}+v^{-, n}_{\rm i});\ v^{n}_{i+\HALF} = \frac{1}{2}(v^{+, n}_{\rm i+1}+v^{-, n}_{\rm i+1});\  p^{n}_{\rm cr, i} = {\rm same};\ \, \ v^{n}_{\rm i} < 0  \nonumber
\end{cases}\nonumber   \\  \nonumber\\
 {\rm  Possibility\, 2:} & & 
\begin{cases}
    v^{n}_{i-\HALF} = \frac{\cU^{n}_{\rm [mx], i - \HALF}}{\cU^{n}_{\rm [\rho], i - \HALF}};\ v^{n}_{i+\HALF} = \frac{\cU^{n}_{\rm [mx], i + \HALF}}{\cU^{n}_{\rm [\rho], i + \HALF}};\, p^{n}_{\rm cr, i} =\frac{1}{2}(p^{+,n}_{\rm cr, i}+p^{-,n}_{\rm cr, i+1});\ \, \ v^{n}_{\rm i}\geq 0 \nonumber \\ 
   v^{n}_{i-\HALF} = {\rm same};\ v^{n}_{i+\HALF} ={\rm same};\  p^{n}_{\rm cr, i} = \frac{1}{2}(p^{+,n}_{\rm cr, i-1}+p^{-,n}_{\rm cr, i});\ \, \ v^{n}_{\rm i} < 0  \nonumber
\end{cases}\nonumber \\  \nonumber\\
{\rm  Possibility\, 3:} & & 
\begin{cases}
    v^{n}_{i-\HALF} = \frac{\cU^{n}_{\rm [mx], i - \HALF}}{\cU^{n}_{\rm [\rho], i - \HALF}};\ v^{n}_{i+\HALF} = \frac{\cU^{n}_{\rm [mx], i + \HALF}}{\cU^{n}_{\rm [\rho], i + \HALF}};\, p^{n}_{\rm cr, i} =\frac{1}{2}(p^{+,n}_{\rm cr, i-1}+p^{-,n}_{\rm cr, i});\ \, \ v^{n}_{\rm i}\geq 0 \nonumber \\ 
   v^{n}_{i-\HALF} = {\rm same};\ v^{n}_{i+\HALF} ={\rm same};\  p^{n}_{\rm cr, i} = \frac{1}{2}(p^{+,n}_{\rm cr, i}+p^{-,n}_{\rm cr, i+1});\ \, \ v^{n}_{\rm i} < 0  \nonumber
\end{cases}\nonumber 
\end{eqnarray} \nonumber
\begin{figure}
\centering
\includegraphics[height= 3.7in,width=6.2in]{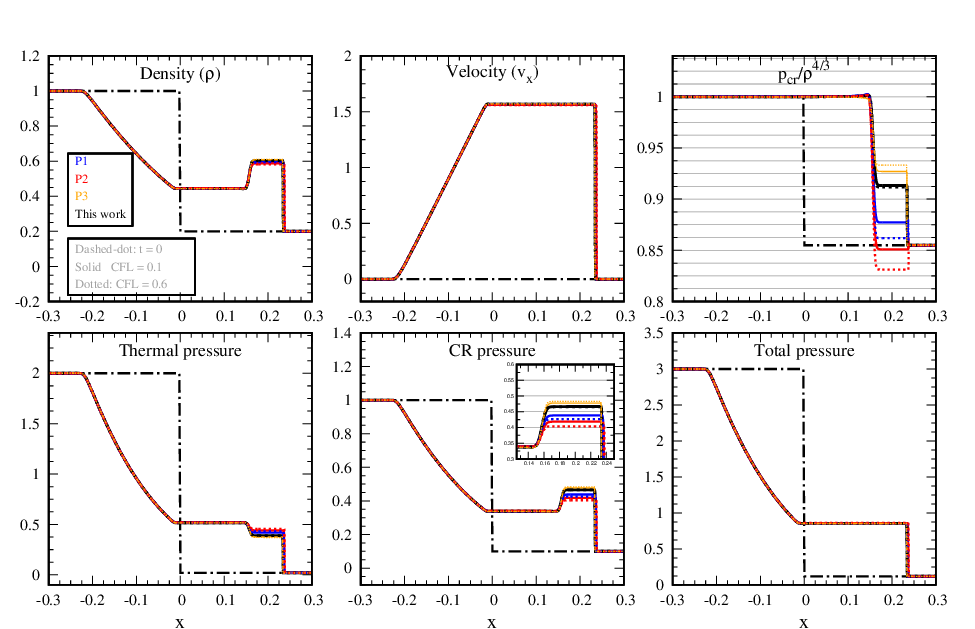}
\caption{Two-fluid shock tube (problem $2$  in Table \ref{tab:testproblems}). The dashed-dotted black curves show profiles at $t=0$ and the solid/dotted curves show profiles at $t=0.1$. For all runs, the used method is \quotes{\emph{Unsplit-pdv}}, however, the choice of $v$ and $p_{\rm cr}$ are different (see the Possibilities $1-3$ described in Appendix \ref{app:B}). The solutions are obtained using a $1^{\rm st}$ order numerical scheme, where $N_{\rm x} = 5000$, but  two different CFL numbers: for solid curves $C_{\rm a} = 0.1$ and dotted curves $C_{\rm a} = 0.6$. The figure shows that the solution in the post-shock region depends not only on the choice of the numerical method but also the CFL number. For our preferred method, shown with solid and dotted black lines, solutions are robust to the choice of CFL number. \BLACK} 
\label{fig:figpdv}
\end{figure}
Note that the possible choices are not limited to the above three combinations. Despite the similarity of these choices, we find that the results are not identical and also depend on the CFL number. Fig. \ref{fig:figpdv} shows the solutions for two different CFL numbers $0.1$ (solid curves) and $0.6$ (dotted curves) respectively. Each colour represents different possible combinations of $p_{\rm cr}$ and $v$ given above. We have tried several other possibilities and found that all give different result. This experiment shows that our choice of $v$ and $p_{\rm cr}$ (i.e., Eqs. \ref{eq:v} and \ref{eq:pcr}) provides a robust result (solid/dotted black curves), i.e., solutions are robust to the choice of spatial reconstructions (Fig. \ref{fig:sodcr1}), time stepping (Fig. \ref{fig:sodcr1zoom}), and the CFL number (Fig. \ref{fig:figpdv}).


\label{lastpage}
\end{document}